\documentclass[aps,prl,twocolumn,superscriptaddress]{revtex4-1} 

\usepackage{graphicx}
\usepackage{dcolumn}
\usepackage{bm}
\usepackage{hyperref}
\usepackage{todonotes}
\usepackage{color}
\usepackage{comment}
\usepackage{sidecap}
\sidecaptionvpos{figure}{c}
\usepackage{lipsum}
\usepackage{bm}
\usepackage{amsmath}

\begin{document}

\def\mathbi#1{\textbf{\em #1}}  
\def\Q{\ensuremath{\mathbi{Q}}}
\def\pipi{\ensuremath{(\pi, \pi)}}

\title{Doping dependence of the magnetic excitations in La$_{2-x}$Sr$_{x}$CuO$_4$}

\author{D. Meyers}
\author{H. Miao}
\affiliation{Department of Condensed Matter Physics and Materials Science, Brookhaven National Laboratory, Upton, New York 11973, USA}
\email{dmeyers@bnl.gov}

\author{A. C. Walters}
\affiliation{Diamond Light Source Ltd., Harwell Science and Innovation Campus, Chilton, Didcot, Oxfordshire, OX11 0DE, UK}

\author{V. Bisogni}
\affiliation{National Synchrotron Light Source II, Brookhaven National Laboratory, Upton, New York 11973, USA}

\author{R. S. Springell}
\affiliation{Interface Analysis Centre, School of Physics, University of Bristol, Bristol, BS2 8BS, UK}

\author{M. d'Astuto}
\affiliation{Institut de Min\'{e}ralogie et de Physique des Milieux Condens\'{e}s (IMPMC),
UMR CNRS 7590, Universit\'{e} Pierre et Marie Curie - case 115,
4, place Jussieu, 75252 Paris cedex 05, France}

\author{M. Dantz}
\author{J. Pelliciari}
\affiliation{Research Department ``Synchrotron Radiation and Nanotechnology'', Paul Scherrer Institut, CH-5232 Villigen PSI, Switzerland}

\author{H. Huang}
\author{J. Okamoto}
\affiliation{National Synchrotron Radiation Research Center, Hsinchu 30076, Taiwan}

\author{D. J. Huang}
\affiliation{National Synchrotron Radiation Research Center, Hsinchu 30076, Taiwan}
\affiliation{Department of Physics, National Tsing Hua University, Hsinchu 30013, Taiwan}

\author{J.P. Hill}
\affiliation{National Synchrotron Light Source II, Brookhaven National Laboratory, Upton, New York 11973, USA}

\author{X. He}
\author{I. Bo\v{z}ovi\'{c}}
\affiliation{Department of Condensed Matter Physics and Materials Science, Brookhaven National Laboratory, Upton, New York 11973, USA}
\affiliation{Yale University, Applied Physics Department, New Haven CT 06520, USA}

\author{T. Schmitt}
\affiliation{Research Department ``Synchrotron Radiation and Nanotechnology'', Paul Scherrer Institut, CH-5232 Villigen PSI, Switzerland}

\author{M. P. M. Dean}
\email{mdean@bnl.gov}
\affiliation{Department of Condensed Matter Physics and Materials Science, Brookhaven National Laboratory, Upton, New York 11973, USA}

\begin{abstract}

The magnetic correlations within the cuprates have undergone intense scrutiny as part of efforts to understand high temperature superconductivity. We explore the evolution of the magnetic correlations along the nodal direction of the Brillouin zone in La$_{2-x}$Sr$_{x}$CuO$_4$, spanning the doping phase diagram from the anti-ferromagnetic Mott insulator at $x = 0$ to the metallic phase at $x = 0.26$. Magnetic excitations along this direction are found to be systematically softened and broadened with doping, at a higher rate than the excitations along the anti-nodal direction. This phenomenology is discussed in terms of the nature of the magnetism in the doped cuprates. Survival of the high energy magnetic excitations, even in the overdoped regime, indicates that these excitations are marginal to pairing, while the influence of the low energy excitations remains ambiguous.

\end{abstract}

\maketitle

\section{Introduction}
\begin{figure}[h]
\includegraphics[width=.52\textwidth]{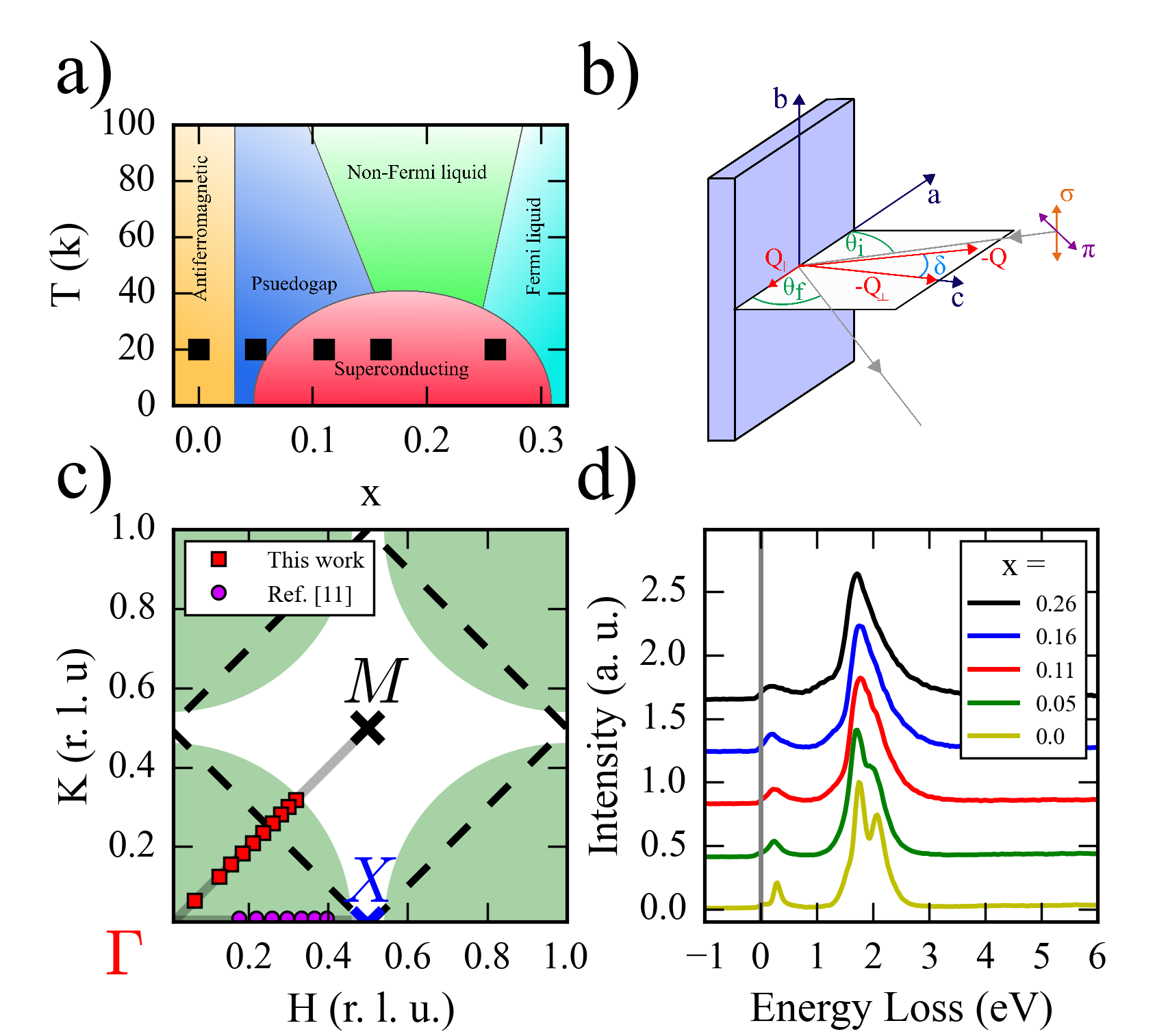} %
\caption{a) Doping phase diagram for LSCO. b) RIXS geometry used for the experiment. c) Plot of the cuprate Brillouin zone. Red squares represent \Q-points measured for this report, purple circles for previous work by this group \cite{DeanLSCO2013}. The green semi-circles represent the regions of reciprocal space accessible to Cu L$_3$-edge RIXS. d) RIXS spectra for each doping at $\Q{} = (0.24, 0.24)$ showing $dd$-excitation and paramagnon features. Data are offset vertically for clarity. The grey line denotes the zero-energy loss position. }
\label{Intro}
\end{figure}

The past several decades have witnessed a considerable scientific effort within the condensed matter community to unravel the true origin of high temperature superconductivity (HTS) \cite{Dagotto1994, Lee2006, Scalapino2012, Keimer2015}. Due to the proximity of antiferromagnetic order and HTS in the doping phase diagram, as shown in Fig.~\ref{Intro}(a), the relationship between magnetism and superconductivity has been discussed extensively. This even includes postulating that superconducting pairing is driven by the exchange of magnetic excitations, although putting such a scenario on firm theoretical footing remains very challenging \cite{Eschrig2006}. What is clear, however, is that the Coulomb repulsion, $U$, plays a dominant role in the formation of the anti-ferromagnetic (AFM) insulating state in the cuprate parent compounds and that this leads to the emergence of well-defined spin wave or magnon excitations \cite{Coldea2001, Braicovich2010, Guarise2010, Headings2010, Dean2012, DeanLSCO2013, Dean2015, Lebert2016, Dantz2016}. These disperse up to very high energies (about 300~meV) along the anti-nodal direction in the Brillouin zone [$(0,0) \rightarrow (0.5, 0)$ in commonly used tetragonal notation] and up to comparable but somewhat lower energies along the nodal [$(0,0) \rightarrow (0.25, 0.25)$] direction. Such a large energy scale would naturally be expected to play a central role over a large fraction of the phase diagram. There have consequently been extensive efforts to characterize the nature of magnetism across the cuprate phase diagram \cite{Bourges2000, Tranquada2004, Hayden2004, Woo2006, Lipscombe2007,Vignolle2007, LeTacon2011, Bisogni2012, DeanLSCO2013, DeanBSCCO2013, DeanBSCCO2014, Guarise2014, Dean2015, Wakimoto2015, Ellis2015, Huang2016,Monney2016}, which have demonstrated that substantial magnetic spectral weight persists to at least optimal doping (i.e.\ a hole concentration $x\approx 0.15$).  A more controversial issue is what happens for the overdoped cuprates with hole concentration $0.15 \lesssim x \lesssim 0.3$. In this doping range inelastic neutron scattering (INS) indicates that the energy, $\omega$, and wavevector, \Q{}, integrated magnetic spectral weight is strongly reduced \cite{Fujita2012NS}. This even encompasses discussions of whether magnetism might effectively disappear completely in the overdoped regime and whether this might be the reason that HTS is suppressed at a similar doping level \cite{Scalapino2012, Wakimoto2004, Fujiyama2012}. More extensive studies of the overdoped cuprates, and how their excitations relate to the magnons in the parent compounds, are therefore desirable. In recent years resonant inelastic x-ray scattering (RIXS) \cite{Ament2011, Dean2015} has emerged as a complementary tool to INS in the investigation of magnetic excitations in the cuprates \cite{LeTacon2011, DeanLSCO2013, DeanBSCCO2013, DeanBSCCO2014, Guarise2014, Dean2015, Wakimoto2015, Huang2016,Monney2016}. In particular, RIXS offers the possibility to probe very small sample volumes \cite{Dean2012} and to perform more extensive doping dependence as the large single crystal samples required for INS are often highly challenging to produce for overdoped cuprates.

In this article, we use RIXS to investigate La$_{2-x}$Sr$_{x}$CuO$_4$ (LSCO) along the nodal direction of the Brillouin zone in samples spanning the AFM insulating to overdoped superconducting phases in order to unravel the doping systematics of the magnetic excitations. Combined with our previous equivalent data along the anti-nodal direction \cite{Dean2012,DeanLSCO2013}, this constitutes an extensive characterization of the high energy magnetic excitation spectrum throughout the LSCO phase diagram. We find a gradual softening and broadening of the paramagnon feature with doping. The high-energy magnetic excitations studied here do not show any strong changes that correlate with the suppression of superconductivity in the overdoped regime, consistent with previous work asserting that these modes have a marginal role in HTS \cite{DeanLSCO2013}. This is notably different from the low-energy AFM correlations around $(0.5,0.5)$ which \emph{are} strongly modified in the overdoped regime. Qualitatively similar differences between the nodal and anti-nodal directions are captured by calculations of the magnetic excitation spectrum based on itinerant quasi-particles, which is discussed in terms of conceptualizing magnetism in the overdoped cuprates \cite{James2012, DeanBSCCO2013, Zeyher2013, Eremin2013, Li2016}.

\section{Experimental methods}
Thin film LSCO samples were synthesized on single-crystal LaSrAlO$_4$ substrates with atomic layer-by-layer molecular beam epitaxy \cite{Bozovic2001}. Typical surface root-mean-square roughnesses, as determined by atomic force microscopy, were about 3~\AA{}, which helps to reduce the contribution of diffuse elastic scattering to the RIXS spectra. The sample thicknesses for $x=0$, 0.05, 0.11, 0.16 and 0.26, were 53, 79, 99, 53 and 99~nm respectively, as determined by measuring the Kiessig fringes in x-ray diffraction, consistent with counting the layers during the growth process. The RIXS data presented were collected using the SAXES spectrometer \cite{Ghiringhelli2006} at the ADRESS beamline \cite{Strocov2010} of the Swiss Light Source at the Paul Scherrer Institute. Further checks were performed at the AGS-AGM spectrometer \cite{Lai2014} at BL05A1 -- the Inelastic Scattering Beamline at the National Synchrotron Radiation Research Center, Taiwan. Energy resolution measurements carried out on carbon tape immediately before sample measurements gave an overall energy resolution of $\sim 120$~meV full width at half maximum (FWHM) and served as a reference for the zero energy loss calibration. All data shown were collected at low temperature (approximately 20~K).

Figure~\ref{Intro}(b) displays the  horizontal scattering geometry used for this study. The momentum transferred in-plane was varied by changing the incident angle, $\theta_{i}$, giving, $\Q_{\parallel} = 2\mathopen|\mathbi{k}_i\mathclose| \sin(2\theta/2) \sin(\delta)$, where $\delta =2\theta/2-\theta_i$. The x-ray scattering angle, $2\theta$, was fixed at 130$^{\circ}$ and measurements were taken with grazing exit geometry ($\theta_i>65^{\circ}$) and horizontal incident x-ray polarization equivalent to previous studies \cite{LeTacon2011, DeanLSCO2013, DeanBSCCO2013, DeanLBCO2013, DeanBSCCO2014}. Figure~\ref{Intro}(c) plots the two dimensional Brillouin zone as a function of $\Q=(H,K)$, which is defined in terms of the high temperature tetragonal unit cell with $a=b\approx 3.8$~\AA{}. Within this study, measurements were taken along the nodal direction (from $\Gamma$  towards $M$), which we consider in the context of previously collected data taken along the anti-nodal direction (from $\Gamma$ towards $X$) \cite{DeanLSCO2013}.

\section{Results and analysis}

\begin{figure}[b]
\includegraphics[width=0.5\textwidth]{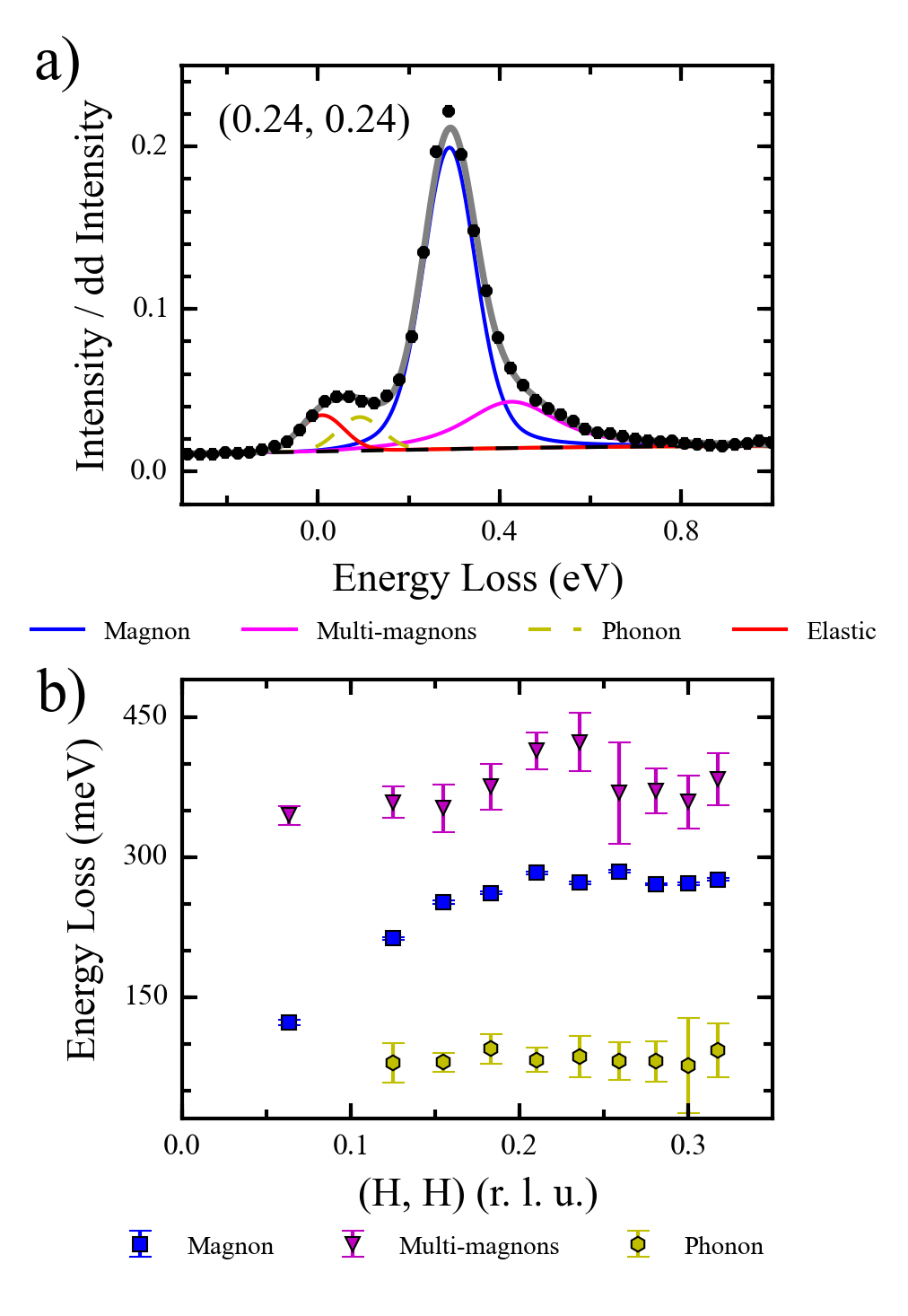} %
\caption{La$_2$CuO$_4$ RIXS spectra. a) An example spectrum at $\Q = (0.24, 0.24)$ showing the different components of the fit including magnon, multi-magnon, phonon and elastic as well as the total fit in gray. b) Dispersion of the phonon, multi-magnon, and single magnon features.}
\label{0Doping}
\end{figure}

Figure~\ref{Intro}(d) shows example RIXS spectra at $\Q=(0.24,0.24)$ as a function of doping. La$_2$CuO$_4$ shows a small elastic feature with sharp magnetic  and well defined $dd$ features consistent with the localized insulating nature of the parent compound. As the doping traverses the antiferromagnetic (AFM), pseudogap, and under- and over-doped superconducting regions the $dd$ features become broadened. 

\begin{figure}
\includegraphics[width=0.5\textwidth]{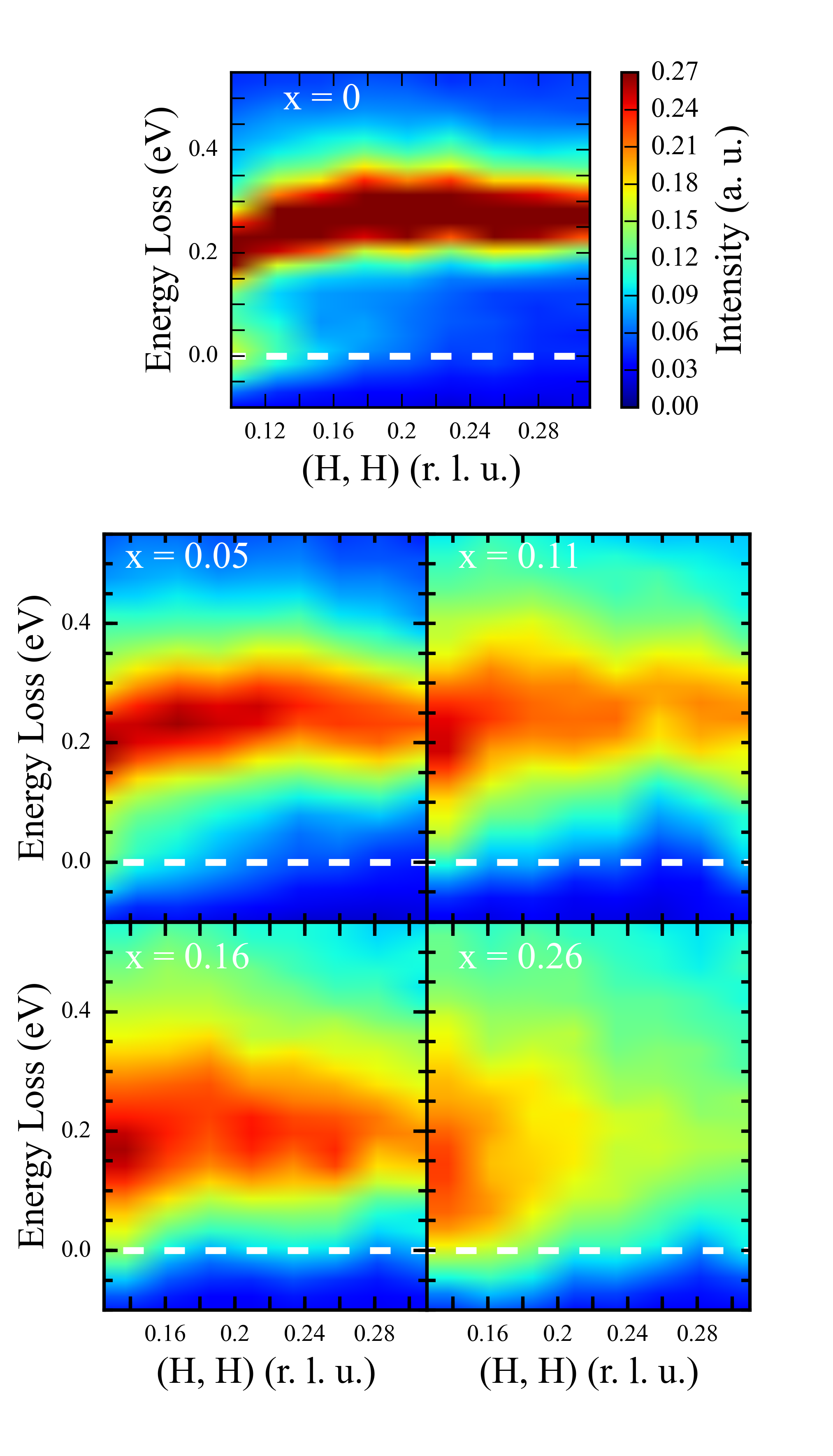} %
\caption{Plots of the RIXS spectra dispersion for each doping represented as a 2D colormap. The broadening of the paramagnon feature with increased doping is readily apparent. All panels share the same intensity scale as shown on the top panel. The doping, $x$, is noted in white in the top left corner of every panel.}
\label{2D_plots}
\end{figure}

Magnetic excitations along the nodal direction in the Brillouin zone in the parent compound La$_2$CuO$_4$ have been studied in detail and can be adequately modeled using spin-wave theory \cite{Coldea2001,Headings2010,Braicovich2010,Dean2012}. In RIXS spectra these spin wave or magnon excitations are present alongside an optical phonon and high-energy magnetic continuum \cite{Braicovich2010,Guarise2010,Dean2012,Dean2015}. We performed a similar analysis to that used  previously in Refs.~\cite{LeTacon2011, DeanLSCO2013, DeanLBCO2013, DeanBSCCO2014, Dean2015} in order to facilitate direct comparisons with previous work. An example fit is shown in Fig.~\ref{0Doping}(a). Here, the magnon and magnetic continuum features were modeled with an anti-symmetrized Lorentzian convolved with a Gaussian resolution function with FWHM $\sim 120$~meV, Fig.~\ref{0Doping}(a) \footnote{See the Supplemental Material of Ref.~\cite{DeanLSCO2013} for a explicit definition of the function}. The anti-symmetrized Lorentzian is used to account for the time reversal symmetry of the imaginary part of the dynamical susceptibility, $\chi^{\prime\prime}(\Q,\omega)$. Following previous work, this is proportional to the scattering function corrected by the Bose factor
\begin{equation}
    S(\Q,\omega) \propto \frac{\chi^{\prime\prime}(\Q,\omega)}{1-\exp(-\omega/k_BT)},
\end{equation}
where $k_B T$ is the thermal energy scale \cite{LeTacon2011,DeanLSCO2013,DeanBSCCO2014}. The elastic scattering and optical phonon were fit with Gaussian functions. Finally, a smooth background was fit with a $3^{\text{rd}}$-order polynomial to account for the tail of the $dd$ excitations that carry much higher intensity, see Fig.~\ref{Intro}(d). This model provides a good description of the spectral lineshape for the $x=0$ insulator below 1~eV. The broader features at higher dopings makes it impossible to convincingly separate the multi-magnon and the phonon features from the paramagnon and we perform fits in which a single peak accounts for all these features to allow for direct comparison with previous work \cite{LeTacon2011,DeanLSCO2013,DeanBSCCO2014}. We also note that a recent report discussed the appropriateness of this widely used scheme in the case of heavily overdamped modes \cite{Lamsal2016}. We examined this issue in detail as described in the Supplemental Material, concluding that the functional form we use is adequate for the damping rates found in this study \cite{supplemental}.

\begin{figure*}
\centering
\includegraphics[width=\textwidth]{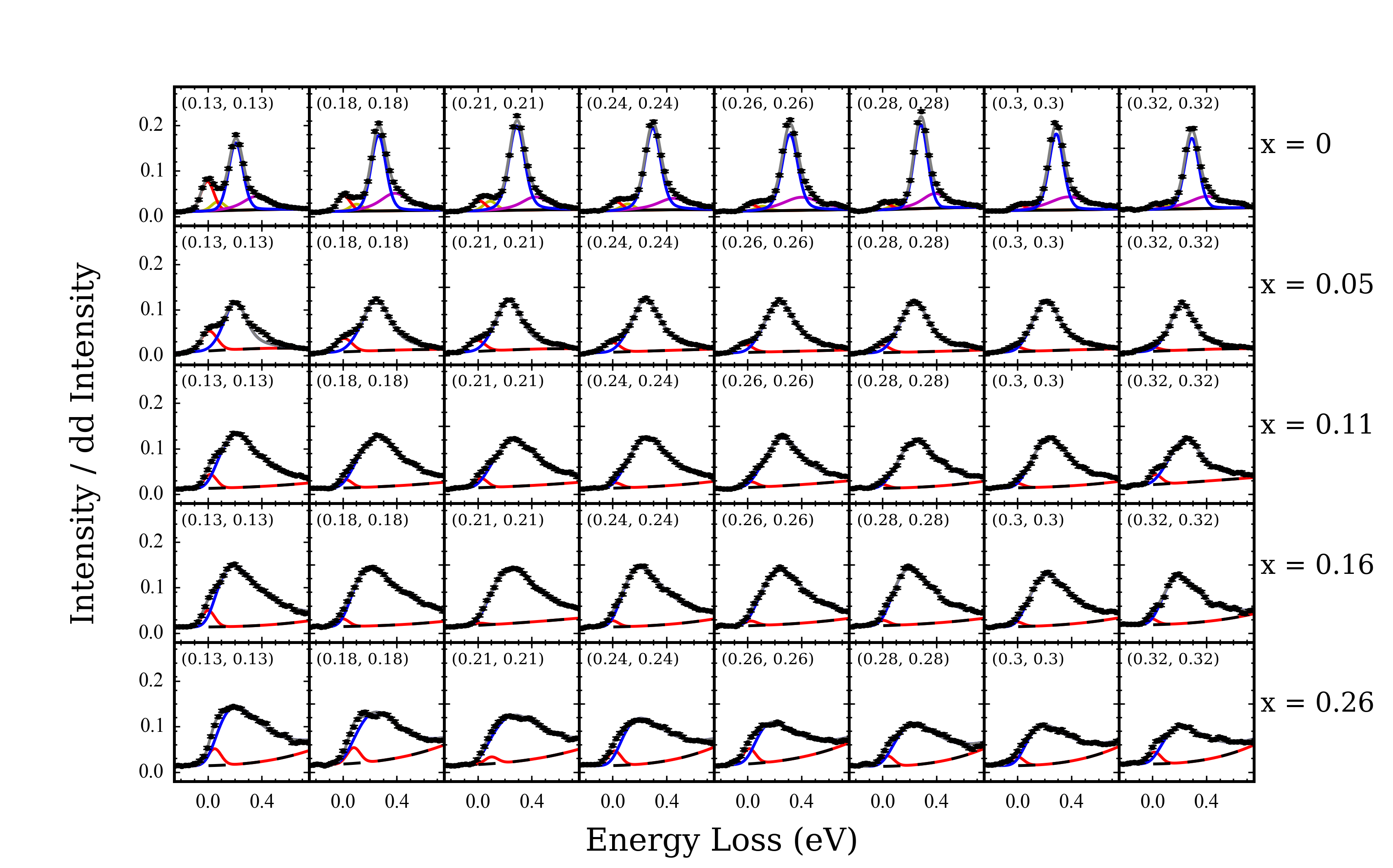} %
\caption{Individual RIXS spectra for each sample (horizontal) and every \Q-point (vertical) with underlain fit. For all samples except the parent compound only the elastic (red), paramagnon (blue), and smooth background (dashed black) were used  as components in the fit. The total fit is represented by a grey line.}
\label{All_Specs}
\end{figure*}

Figure \ref{0Doping}(b) shows the fitting results for La$_2$CuO$_4$ $x=0$. A dispersive magnon is observed with a maximum energy transfer of $\sim290$~meV around the Brillouin zone boundary consistent with spin wave theory predictions based on INS \cite{Coldea2001,Headings2010} and previous RIXS studies \cite{Dean2012}. The magnetic continuum feature is challenging to unambiguously separate from the magnon, but, as expected, it lies at higher energies and is less dispersive than the magnon. The 90~meV feature has negligible \Q{}-dependence ($\lesssim 25$~meV), limited by the errorbars, consistent with it being an optical phonon \cite{Padilla2005,Cohen1990, Wang1999}.

\begin{figure*}
\centering
\includegraphics[width=1\textwidth]{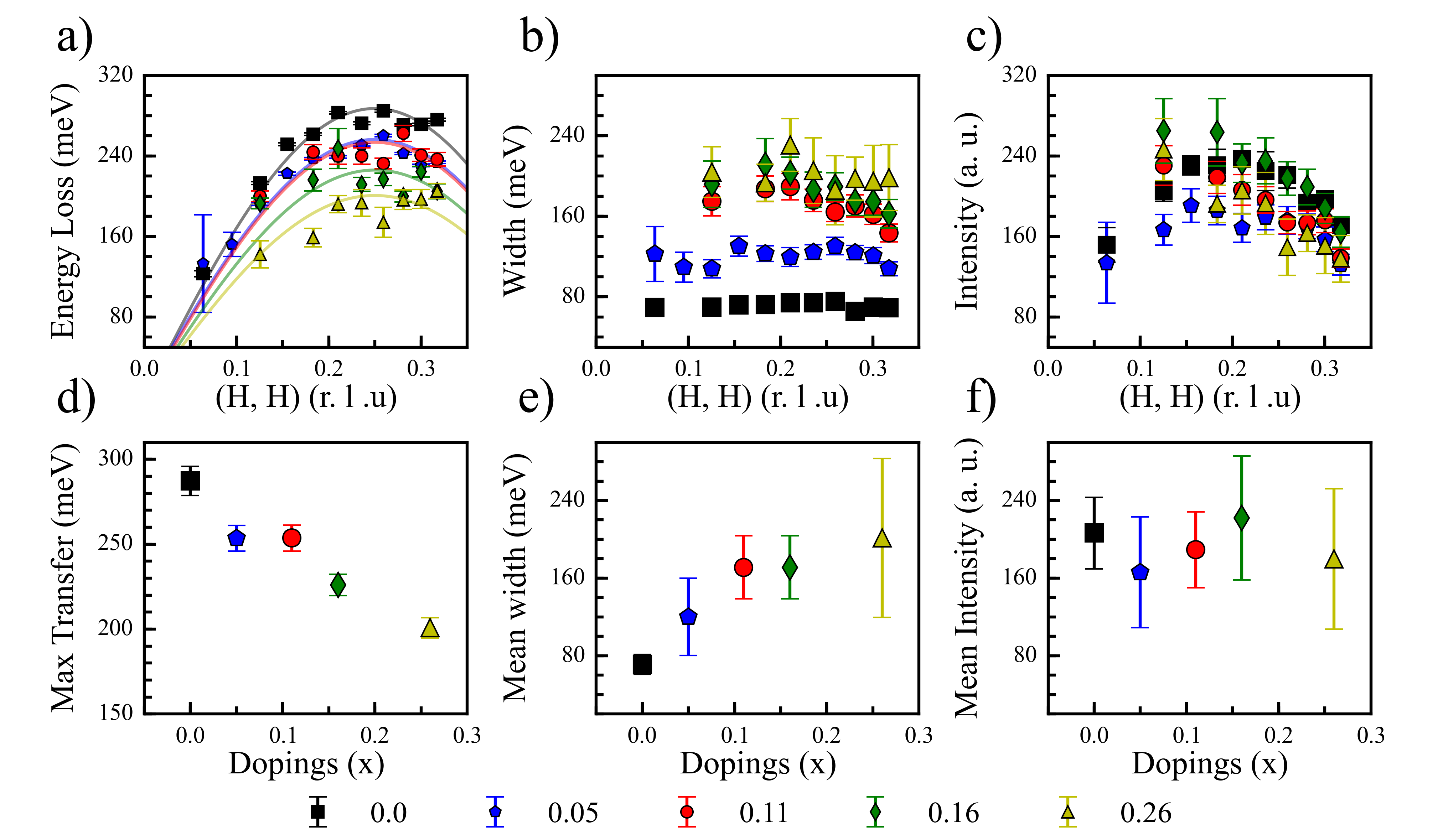} %
\caption{ a) The paramagnon dispersion for each doping with applied fit. b) Paramagnon peak width for each \Q-point defined as Lorentizian half-width at half-maximum (HWHM).  c) Paramagnon spectral weight for each sample and \Q-point.  d) Maximum energy transfer for each doping. e) \Q-averaged widths with cumulative error. f)  \Q-averaged intensities. The error bars represent the statistical error from the fitting procedure.}
\label{Dis_Wid_Int}
\end{figure*}

We now discuss the doping evolution of the nodal excitations plotted as a colormap in Fig.~\ref{2D_plots}. Low energy spectral features  are visible for all studied dopings. Detailed incident energy dependence has confirmed the magnetic nature of these spectral features \cite{Minola2015, Huang2016}, consistent with theoretical calculations \cite{Jia2014} and the smooth evolution of these features with respect to the unambiguously magnetic feature in La$_2$CuO$_4$. Future experiments that explicitly resolve the scattered x-ray polarization would be a valuable addition to this issue \cite{Braicovich2014}. In comparison to the parent compound, the magnetic spectral weight is broadened and appears to be softened. Both these effects appear to occur continuously as the doping spans the phase diagram from the pseudogap  ($x = 0.05$), underdoped ($x = 0.11$), nearly optimal ($x = 0.16$), and all the way to the overdoped phase ($x = 0.26$). This phenomenology is consistent with previous examinations of the nodal dispersion in Bi$_2$Sr$_2$Ca$_{n-1}$Cu$_n$O$_{2n+4+\delta}$ \cite{Guarise2014, DeanBSCCO2014, Huang2016} and LSCO ($x = 0.26$, $0.30$) \cite{Wakimoto2015, Monney2016}. The present data, alongside Ref.~\cite{DeanLSCO2013}, provide the first comprehensive RIXS study of undoped, underdoped, optimally doped and overdoped samples within a single cuprate system. The characteristic energy of the excitations was addressed in more detail by fitting analysis as shown in Fig.~\ref{All_Specs}.  As can be seen, the fitting procedure is able to reliably reproduce the experimental curves despite the significant broadening at higher dopings. 

\begin{figure}
\includegraphics[width=0.5\textwidth]{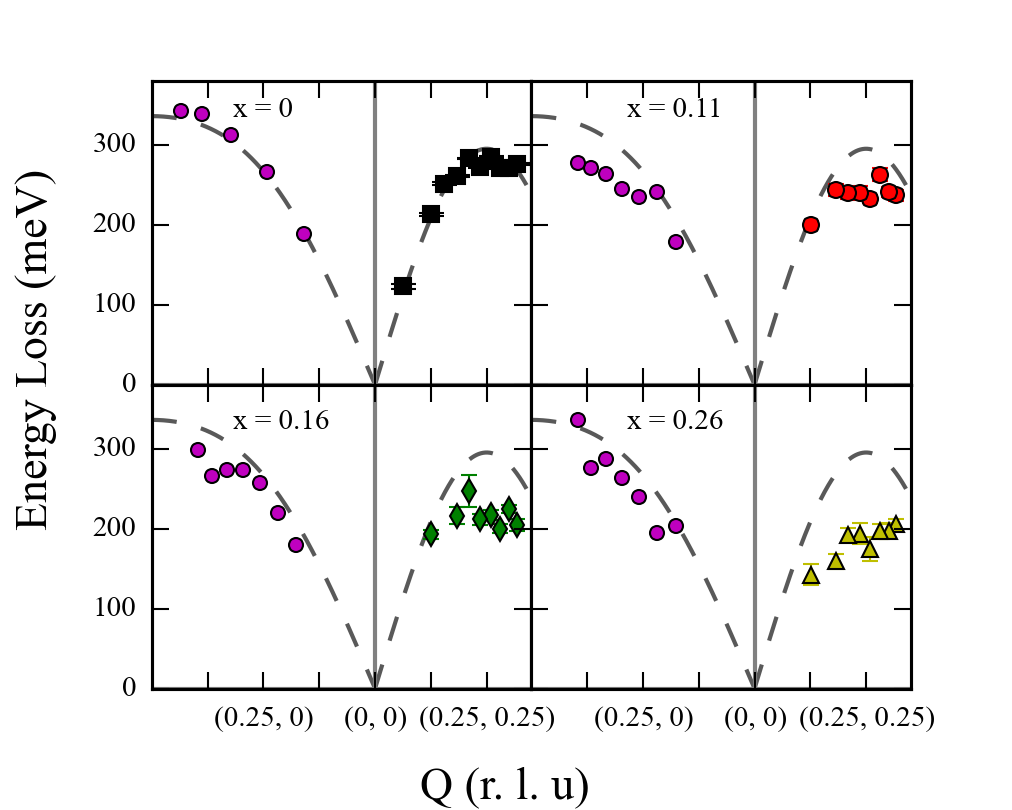} %
\caption{Comparison of the nodal and anti-nodal \cite{DeanLSCO2013,Dean2012} data for each doping. Dashed lines represent the spin-wave theory predictions for La$_2$CuO$_4$. x = 0.05 was not previously measured.}
\label{RecipSpace}
\end{figure}

 Figure~\ref{Dis_Wid_Int} provides the values from the least squares fitting analysis.  A softening of the paramagnon mode with doping in Fig.~\ref{Dis_Wid_Int}(a) is very apparent, with the overdoped sample in particular showing a significant drop of around $80\pm10$~meV, with respect to $x=0$, at high \Q{}. The paramagnon energy also becomes progressively less \Q{}-dependent with doping. We extracted a characteristic energy scale for the excitations by fitting a sinusoidal function \cite{Coldea2001},  Figure~\ref{RecipSpace},  to the \Q{}-dependence at different $x$ and plot the results in Fig.~\ref{Dis_Wid_Int}(d) \footnote{The undoped dispersion function with fixed magnetic exchange and a variable maximum energy transfer was used for the fitting.}. The maximum energy transfer follows a nearly linear dependence upon doping, irrespective of the notable changes in the electronic state of the sample (i.e.\ insulating, pseudogap, superconducting). Moving to the paramagnon width [Fig.~\ref{Dis_Wid_Int}(b)], fitting the $x=0$ spectrum  yields a value of around 75~meV, which is likely to come from difficulties separating contributions from phonons and the magnetic continuum from the magnetic pole \cite{DeanLSCO2013} and should be thought of as an upper limit on the width. Upon doping the average width [Fig.~\ref{Dis_Wid_Int}(e)] increases strongly. The integrated spectral intensity, normalized to the $dd$ intensity, [Fig.~\ref{Dis_Wid_Int}(c)] peaks at intermediate \Q{} values and persists without strong changes as a function of doping\footnote{For the x = 0 case, the phonon and multi-magnon intensities were included with the paramagnon intensity as these were not distinguishable for the higher dopings. }. Self-absorption can satisfactorily explain the decrease towards higher \Q, as the x-ray emission angle is progressively lowered. Figure~\ref{Dis_Wid_Int}(e)\&(f) show that the spectra broaden continuously with increasing doping without strong changes in integrated spectral weight.

\section{Discussion}

A comparison between the nodal and anti-nodal dispersions, and how they relate to the dispersion in the parent compound, is presented in Fig.~\ref{RecipSpace}. The dashed line represents the dispersion from spin-wave theory for La$_2$CuO$_4$ \cite{Coldea2001}. Both the nodal and anti-nodal directions are shown, with the results for the same samples from a previous work displayed as purple circles \cite{Dean2012, DeanLSCO2013}. Dispersive magnetic excitations are clearly present for all dopings, albeit with strongly reduced dispersion particularly in the nodal direction. This extends the discussion of the behavior of the magnetic excitations in the anti-nodal direction \cite{DeanLSCO2013} to the nodal direction within the Brillouin zone. It is clear that the overdoped cuprates do not become trivial, completely non-magnetic, Fermi-liquids. Rather the effects of $U$ still appear to be present in the form of strongly damped high energy magnetic excitations within the region in reciprocal space we have studied [shown in Fig.\ref{Intro}(c)].

These results help constrain theoretical models of magnetism in the cuprates. The Hubbard model, in either its one or three band variants, is often considered to be the best-justified starting point for modeling magnetism in the cuprates (and strongly correlated materials more generally) \cite{Lee2006}. Calculations of the Hubbard model indeed capture the persistent high energy magnetic excitations in good accord with experiments \cite{Macridin2007,Jia2014}. However, such calculations are typically confined to small lattices and there is significant interest in conceptualizing magnetism in terms of either local moment \cite{Vojta2009, Xu2009} or (renormalized) itinerant quasiparticle methods \cite{James2012, DeanBSCCO2013, Zeyher2013, Eremin2013, Li2016}. A local moment based approach is well justified in La$_2$CuO$_4$ and has been used a lot in terms of describing stripe physics in the underdoped regime \cite{Vojta2009}. Such an approach also naturally explains why the intensity of the excitations in the doped cuprates is of a comparable magnitude to those in the insulator. Itinerant quasi-particle pictures, in which correlations are treated within several different approximate schemes, are also popular. These have a significant advantage in terms of allowing predictions with high \Q{}-resolution \cite{James2012, DeanBSCCO2013, Zeyher2013, Eremin2013, Li2016, Monney2016}. They also capture the fact that the nodal excitations are more strongly doping dependent than the anti-nodal excitations, as observed here. 

It is also important to consider the data presented here in the context of previous INS measurements \cite{Coldea2001, Headings2010, Bourges2000, Tranquada2004, Hayden2004, Woo2006, Lipscombe2007, Vignolle2007, Wakimoto2015, Dean2015}. Spin wave theory for AFM La$_2$CuO$_4$ implies that the dispersion around $(0,0)$ and $(0.5,0.5)$ should be symmetric. On this basis, it is tempting to compare RIXS and INS data on doped cuprates assuming this symmetry. Figure~\ref{RecipSpace} shows that for $x=0.26$ the magnetic excitations at wavevectors accessed by RIXS around $(0,0)$ have a significantly lower energy scale than excitations seen by neutrons around $(0.5,0.5)$ \cite{Wakimoto2015} suggesting that this symmetry breaks down in the doped cuprates and such a comparison appears invalid. Unfortunately, there is very little directly comparable RIXS and INS data taken at the same \Q{}, however, the one available study suggests that both methods provide similar access to magnetic dispersions \cite{Wakimoto2015}.

In the context of HTS, we note that the high energy excitations studied here show minimal changes in going from the optimally doped to the overdoped regime. As such, they further support previous assertions that the high-energy magnetic excitations observed here are marginal to superconducting pairing \cite{DeanLSCO2013}. This hypothesis tallies with a simple consideration of how a repulsive interaction might contribute to $d$-wave pairing in the cuprates via transitions between different states near the Fermi level. In such a scenario the excitations around $(0.5,0.5)$ would be expected to have a stronger contribution than excitations elsewhere in \Q{} and excitations far from $(0.5,0.5)$  can even lead to pair breaking and reduction of $T_c$  \cite{Scalapino2012}.

\section{Conclusions}
The paramagnon dispersion along the nodal direction in the Brillouin zone is reported for samples spanning the LSCO phase diagram from the AFM insulator to the overdoped superconductor. We found a gradual softening and damping of the paramagnon excitation with doping as predicted by Hubbard model calculations \cite{Macridin2007, Jia2014}. The character of the excitations indicates that magnetic correlations in the cuprates have both localized and itinerant character. These findings further support suggestions that the high energy magnetic excitations studied here have a marginal role for HTS \cite{DeanLSCO2013}. Likely, the destruction of other pairing interactions, such as the low energy magnetic excitations, is instead culpable for the drop in $T_c$ with overdoping.

\section{Acknowledgements}
\begin{acknowledgments}
This material is based upon work supported by the U.S.\ Department of Energy, Office of Basic Energy Sciences, Early Career Award Program under Award Number 1047478. J.P.\  and  T.S.\  acknowledge  financial  support  through  the  Dysenos  AG  by  Kabelwerke Brugg AG Holding, Fachhochschule Nordwestschweiz, and the Paul Scherrer Institut. M.D.\ and T.S.\ acknowledge funding from the Swiss National Science Foundation within the D-A-CH programme (SNSF Research Grant 200021L 141325). MBE synthesis (I.B.\ and H.X.) was supported by the U.S. Department of Energy, Basic Energy Sciences, Materials Sciences and Engineering Division. Experiments were performed at the ADRESS beam line of the Swiss Light Source at the Paul Scherrer Institut, Switzerland and at BL05A1 - the Inelastic Scattering Beamline at the National Synchrotron Radiation Research Center, Taiwan. Part of this research has been funded by the Swiss National Science Foundation through  the D-A-CH program (SNSF Research Grant No. 200021L 141325).
\end{acknowledgments}


\begin{thebibliography}{56}%
\makeatletter
\providecommand \@ifxundefined [1]{%
 \@ifx{#1\undefined}
}%
\providecommand \@ifnum [1]{%
 \ifnum #1\expandafter \@firstoftwo
 \else \expandafter \@secondoftwo
 \fi
}%
\providecommand \@ifx [1]{%
 \ifx #1\expandafter \@firstoftwo
 \else \expandafter \@secondoftwo
 \fi
}%
\providecommand \natexlab [1]{#1}%
\providecommand \enquote  [1]{``#1''}%
\providecommand \bibnamefont  [1]{#1}%
\providecommand \bibfnamefont [1]{#1}%
\providecommand \citenamefont [1]{#1}%
\providecommand \href@noop [0]{\@secondoftwo}%
\providecommand \href [0]{\begingroup \@sanitize@url \@href}%
\providecommand \@href[1]{\@@startlink{#1}\@@href}%
\providecommand \@@href[1]{\endgroup#1\@@endlink}%
\providecommand \@sanitize@url [0]{\catcode `\\12\catcode `\$12\catcode
  `\&12\catcode `\#12\catcode `\^12\catcode `\_12\catcode `\%12\relax}%
\providecommand \@@startlink[1]{}%
\providecommand \@@endlink[0]{}%
\providecommand \url  [0]{\begingroup\@sanitize@url \@url }%
\providecommand \@url [1]{\endgroup\@href {#1}{\urlprefix }}%
\providecommand \urlprefix  [0]{URL }%
\providecommand \Eprint [0]{\href }%
\providecommand \doibase [0]{http://dx.doi.org/}%
\providecommand \selectlanguage [0]{\@gobble}%
\providecommand \bibinfo  [0]{\@secondoftwo}%
\providecommand \bibfield  [0]{\@secondoftwo}%
\providecommand \translation [1]{[#1]}%
\providecommand \BibitemOpen [0]{}%
\providecommand \bibitemStop [0]{}%
\providecommand \bibitemNoStop [0]{.\EOS\space}%
\providecommand \EOS [0]{\spacefactor3000\relax}%
\providecommand \BibitemShut  [1]{\csname bibitem#1\endcsname}%
\let\auto@bib@innerbib\@empty
\bibitem [{\citenamefont {Dagotto}(1994)}]{Dagotto1994}%
  \BibitemOpen
  \bibfield  {author} {\bibinfo {author} {\bibfnamefont {E.}~\bibnamefont
  {Dagotto}},\ }\href {\doibase 10.1103/RevModPhys.66.763} {\bibfield
  {journal} {\bibinfo  {journal} {Rev. Mod. Phys.}\ }\textbf {\bibinfo {volume}
  {66}},\ \bibinfo {pages} {763} (\bibinfo {year} {1994})}\BibitemShut
  {NoStop}%
\bibitem [{\citenamefont {Lee}\ \emph {et~al.}(2006)\citenamefont {Lee},
  \citenamefont {Nagaosa},\ and\ \citenamefont {Wen}}]{Lee2006}%
  \BibitemOpen
  \bibfield  {author} {\bibinfo {author} {\bibfnamefont {P.~A.}\ \bibnamefont
  {Lee}}, \bibinfo {author} {\bibfnamefont {N.}~\bibnamefont {Nagaosa}}, \ and\
  \bibinfo {author} {\bibfnamefont {X.-G.}\ \bibnamefont {Wen}},\ }\href
  {\doibase 10.1103/RevModPhys.78.17} {\bibfield  {journal} {\bibinfo
  {journal} {Rev. Mod. Phys.}\ }\textbf {\bibinfo {volume} {78}},\ \bibinfo
  {pages} {17} (\bibinfo {year} {2006})}\BibitemShut {NoStop}%
\bibitem [{\citenamefont {Scalapino}(2012)}]{Scalapino2012}%
  \BibitemOpen
  \bibfield  {author} {\bibinfo {author} {\bibfnamefont {D.~J.}\ \bibnamefont
  {Scalapino}},\ }\href {\doibase 10.1103/RevModPhys.84.1383} {\bibfield
  {journal} {\bibinfo  {journal} {Rev. Mod. Phys.}\ }\textbf {\bibinfo {volume}
  {84}},\ \bibinfo {pages} {1383} (\bibinfo {year} {2012})}\BibitemShut
  {NoStop}%
\bibitem [{\citenamefont {Keimer}\ \emph {et~al.}(2015)\citenamefont {Keimer},
  \citenamefont {Kivelson}, \citenamefont {Norman}, \citenamefont {Uchida},\
  and\ \citenamefont {Zaanen}}]{Keimer2015}%
  \BibitemOpen
  \bibfield  {author} {\bibinfo {author} {\bibfnamefont {B.}~\bibnamefont
  {Keimer}}, \bibinfo {author} {\bibfnamefont {S.}~\bibnamefont {Kivelson}},
  \bibinfo {author} {\bibfnamefont {M.}~\bibnamefont {Norman}}, \bibinfo
  {author} {\bibfnamefont {S.}~\bibnamefont {Uchida}}, \ and\ \bibinfo {author}
  {\bibfnamefont {J.}~\bibnamefont {Zaanen}},\ }\href@noop {} {\bibfield
  {journal} {\bibinfo  {journal} {Nature}\ }\textbf {\bibinfo {volume} {518}},\
  \bibinfo {pages} {179} (\bibinfo {year} {2015})}\BibitemShut {NoStop}%
\bibitem [{\citenamefont {Eschrig}(2006)}]{Eschrig2006}%
  \BibitemOpen
  \bibfield  {author} {\bibinfo {author} {\bibfnamefont {M.}~\bibnamefont
  {Eschrig}},\ }\href {\doibase 10.1080/00018730600645636} {\bibfield
  {journal} {\bibinfo  {journal} {Advances in Physics}\ }\textbf {\bibinfo
  {volume} {55}},\ \bibinfo {pages} {47} (\bibinfo {year} {2006})}\BibitemShut
  {NoStop}%
\bibitem [{\citenamefont {Coldea}\ \emph {et~al.}(2001)\citenamefont {Coldea},
  \citenamefont {Hayden}, \citenamefont {Aeppli}, \citenamefont {Perring},
  \citenamefont {Frost}, \citenamefont {Mason}, \citenamefont {Cheong},\ and\
  \citenamefont {Fisk}}]{Coldea2001}%
  \BibitemOpen
  \bibfield  {author} {\bibinfo {author} {\bibfnamefont {R.}~\bibnamefont
  {Coldea}}, \bibinfo {author} {\bibfnamefont {S.~M.}\ \bibnamefont {Hayden}},
  \bibinfo {author} {\bibfnamefont {G.}~\bibnamefont {Aeppli}}, \bibinfo
  {author} {\bibfnamefont {T.~G.}\ \bibnamefont {Perring}}, \bibinfo {author}
  {\bibfnamefont {C.~D.}\ \bibnamefont {Frost}}, \bibinfo {author}
  {\bibfnamefont {T.~E.}\ \bibnamefont {Mason}}, \bibinfo {author}
  {\bibfnamefont {S.-W.}\ \bibnamefont {Cheong}}, \ and\ \bibinfo {author}
  {\bibfnamefont {Z.}~\bibnamefont {Fisk}},\ }\href {\doibase
  10.1103/PhysRevLett.86.5377} {\bibfield  {journal} {\bibinfo  {journal}
  {Phys. Rev. Lett.}\ }\textbf {\bibinfo {volume} {86}},\ \bibinfo {pages}
  {5377} (\bibinfo {year} {2001})}\BibitemShut {NoStop}%
\bibitem [{\citenamefont {Braicovich}\ \emph {et~al.}(2010)\citenamefont
  {Braicovich}, \citenamefont {van~den Brink}, \citenamefont {Bisogni},
  \citenamefont {Sala}, \citenamefont {Ament}, \citenamefont {Brookes},
  \citenamefont {{Luca}}, \citenamefont {Salluzzo}, \citenamefont {Schmitt},
  \citenamefont {Strocov},\ and\ \citenamefont
  {Ghiringhelli}}]{Braicovich2010}%
  \BibitemOpen
  \bibfield  {author} {\bibinfo {author} {\bibfnamefont {L.}~\bibnamefont
  {Braicovich}}, \bibinfo {author} {\bibfnamefont {J.}~\bibnamefont {van~den
  Brink}}, \bibinfo {author} {\bibfnamefont {V.}~\bibnamefont {Bisogni}},
  \bibinfo {author} {\bibfnamefont {M.~M.}\ \bibnamefont {Sala}}, \bibinfo
  {author} {\bibfnamefont {L.~J.~P.}\ \bibnamefont {Ament}}, \bibinfo {author}
  {\bibfnamefont {N.~B.}\ \bibnamefont {Brookes}}, \bibinfo {author}
  {\bibfnamefont {G.~M.~De}\ \bibnamefont {{Luca}}}, \bibinfo {author}
  {\bibfnamefont {M.}~\bibnamefont {Salluzzo}}, \bibinfo {author}
  {\bibfnamefont {T.}~\bibnamefont {Schmitt}}, \bibinfo {author} {\bibfnamefont
  {V.~N.}\ \bibnamefont {Strocov}}, \ and\ \bibinfo {author} {\bibfnamefont
  {G.}~\bibnamefont {Ghiringhelli}},\ }\href {\doibase
  10.1103/PhysRevLett.104.077002} {\bibfield  {journal} {\bibinfo  {journal}
  {Phys. Rev. Lett.}\ }\textbf {\bibinfo {volume} {104}},\ \bibinfo {pages}
  {077002} (\bibinfo {year} {2010})}\BibitemShut {NoStop}%
\bibitem [{\citenamefont {Guarise}\ \emph {et~al.}(2010)\citenamefont
  {Guarise}, \citenamefont {{Dalla Piazza}}, \citenamefont {{Moretti Sala}},
  \citenamefont {Ghiringhelli}, \citenamefont {Braicovich}, \citenamefont
  {Berger}, \citenamefont {Hancock}, \citenamefont {van~der Marel},
  \citenamefont {Schmitt}, \citenamefont {Strocov}, \citenamefont {Ament},
  \citenamefont {van~den Brink}, \citenamefont {Lin}, \citenamefont {Xu},
  \citenamefont {R\o{}nnow},\ and\ \citenamefont {Grioni}}]{Guarise2010}%
  \BibitemOpen
  \bibfield  {author} {\bibinfo {author} {\bibfnamefont {M.}~\bibnamefont
  {Guarise}}, \bibinfo {author} {\bibfnamefont {B.}~\bibnamefont {{Dalla
  Piazza}}}, \bibinfo {author} {\bibfnamefont {M.}~\bibnamefont {{Moretti
  Sala}}}, \bibinfo {author} {\bibfnamefont {G.}~\bibnamefont {Ghiringhelli}},
  \bibinfo {author} {\bibfnamefont {L.}~\bibnamefont {Braicovich}}, \bibinfo
  {author} {\bibfnamefont {H.}~\bibnamefont {Berger}}, \bibinfo {author}
  {\bibfnamefont {J. N.}~\bibnamefont {Hancock}}, \bibinfo {author} {\bibfnamefont
  {D.}~\bibnamefont {van~der Marel}}, \bibinfo {author} {\bibfnamefont
  {T.}~\bibnamefont {Schmitt}}, \bibinfo {author} {\bibfnamefont
  {V.~N.}~\bibnamefont {Strocov}}, \bibinfo {author} {\bibfnamefont
  {L.~J.~P.}~\bibnamefont {Ament}}, \bibinfo {author} {\bibfnamefont
  {J.}~\bibnamefont {van~den Brink}}, \bibinfo {author} {\bibfnamefont {P.-H.}\
  \bibnamefont {Lin}}, \bibinfo {author} {\bibfnamefont {P.}~\bibnamefont
  {Xu}}, \bibinfo {author} {\bibfnamefont {H.~M.}~\bibnamefont {R\o{}nnow}}, \
  and\ \bibinfo {author} {\bibfnamefont {M.}~\bibnamefont {Grioni}},\ }\href
  {\doibase 10.1103/PhysRevLett.105.157006} {\bibfield  {journal} {\bibinfo
  {journal} {Phys. Rev. Lett.}\ }\textbf {\bibinfo {volume} {105}},\ \bibinfo
  {pages} {157006} (\bibinfo {year} {2010})}\BibitemShut {NoStop}%
\bibitem [{\citenamefont {Headings}\ \emph {et~al.}(2010)\citenamefont
  {Headings}, \citenamefont {Hayden}, \citenamefont {Coldea},\ and\
  \citenamefont {Perring}}]{Headings2010}%
  \BibitemOpen
  \bibfield  {author} {\bibinfo {author} {\bibfnamefont {N.~S.}\ \bibnamefont
  {Headings}}, \bibinfo {author} {\bibfnamefont {S.~M.}\ \bibnamefont
  {Hayden}}, \bibinfo {author} {\bibfnamefont {R.}~\bibnamefont {Coldea}}, \
  and\ \bibinfo {author} {\bibfnamefont {T.~G.}\ \bibnamefont {Perring}},\
  }\href {\doibase 10.1103/PhysRevLett.105.247001} {\bibfield  {journal}
  {\bibinfo  {journal} {Phys. Rev. Lett.}\ }\textbf {\bibinfo {volume} {105}},\
  \bibinfo {pages} {247001} (\bibinfo {year} {2010})}\BibitemShut {NoStop}%
\bibitem [{\citenamefont {Dean}\ \emph {et~al.}(2012)\citenamefont {Dean},
  \citenamefont {Springell}, \citenamefont {Monney}, \citenamefont {Zhou},
  \citenamefont {Pereiro}, \citenamefont {Bo{\v z}ovi{\'c}}, \citenamefont
  {Dalla~Piazza}, \citenamefont {R{\o}nnow}, \citenamefont {Morenzoni},
  \citenamefont {van~den Brink}, \citenamefont {Schmitt},\ and\ \citenamefont
  {Hill}}]{Dean2012}%
  \BibitemOpen
  \bibfield  {author} {\bibinfo {author} {\bibfnamefont {M.~P.~M.}\
  \bibnamefont {Dean}}, \bibinfo {author} {\bibfnamefont {R.~S.}\ \bibnamefont
  {Springell}}, \bibinfo {author} {\bibfnamefont {C.}~\bibnamefont {Monney}},
  \bibinfo {author} {\bibfnamefont {K.~J.}\ \bibnamefont {Zhou}}, \bibinfo
  {author} {\bibfnamefont {J.}~\bibnamefont {Pereiro}}, \bibinfo {author}
  {\bibfnamefont {I.}~\bibnamefont {Bo{\v z}ovi{\'c}}}, \bibinfo {author}
  {\bibfnamefont {B.}~\bibnamefont {Dalla~Piazza}}, \bibinfo {author}
  {\bibfnamefont {H.~M.}\ \bibnamefont {R{\o}nnow}}, \bibinfo {author}
  {\bibfnamefont {E.}~\bibnamefont {Morenzoni}}, \bibinfo {author}
  {\bibfnamefont {J.}~\bibnamefont {van~den Brink}}, \bibinfo {author}
  {\bibfnamefont {T.}~\bibnamefont {Schmitt}}, \ and\ \bibinfo {author}
  {\bibfnamefont {J.~P.}\ \bibnamefont {Hill}},\ }\href
  {http://dx.doi.org/10.1038/nmat3409} {\bibfield  {journal} {\bibinfo
  {journal} {Nat. Mater.}\ }\textbf {\bibinfo {volume} {11}},\ \bibinfo {pages}
  {850} (\bibinfo {year} {2012})}\BibitemShut {NoStop}%
\bibitem [{\citenamefont {Dean}\ \emph
  {et~al.}(2013{\natexlab{a}})\citenamefont {Dean}, \citenamefont {Dellea},
  \citenamefont {Springell}, \citenamefont {Yakhou-Harris}, \citenamefont
  {Kummer}, \citenamefont {Brookes}, \citenamefont {Liu}, \citenamefont {Sun},
  \citenamefont {Strle}, \citenamefont {Schmitt}, \citenamefont {Braicovich},
  \citenamefont {Ghiringhelli}, \citenamefont {Bozovic},\ and\ \citenamefont
  {Hill}}]{DeanLSCO2013}%
  \BibitemOpen
  \bibfield  {author} {\bibinfo {author} {\bibfnamefont {M.~P.~M.}\
  \bibnamefont {Dean}}, \bibinfo {author} {\bibfnamefont {G.}~\bibnamefont
  {Dellea}}, \bibinfo {author} {\bibfnamefont {R.~S.}\ \bibnamefont
  {Springell}}, \bibinfo {author} {\bibfnamefont {F.}~\bibnamefont
  {Yakhou-Harris}}, \bibinfo {author} {\bibfnamefont {K.}~\bibnamefont
  {Kummer}}, \bibinfo {author} {\bibfnamefont {N.~B.}\ \bibnamefont {Brookes}},
  \bibinfo {author} {\bibfnamefont {X.}~\bibnamefont {Liu}}, \bibinfo {author}
  {\bibfnamefont {Y.-J.}\ \bibnamefont {Sun}}, \bibinfo {author} {\bibfnamefont
  {J.}~\bibnamefont {Strle}}, \bibinfo {author} {\bibfnamefont
  {T.}~\bibnamefont {Schmitt}}, \bibinfo {author} {\bibfnamefont
  {L.}~\bibnamefont {Braicovich}}, \bibinfo {author} {\bibfnamefont
  {G.}~\bibnamefont {Ghiringhelli}}, \bibinfo {author} {\bibfnamefont
  {I.}~\bibnamefont {Bozovic}}, \ and\ \bibinfo {author} {\bibfnamefont
  {J.~P.}\ \bibnamefont {Hill}},\ }\href {\doibase 10.1038/NMAT3723} {\bibfield
   {journal} {\bibinfo  {journal} {Nature Materials}\ }\textbf {\bibinfo
  {volume} {12}},\ \bibinfo {pages} {1018} (\bibinfo {year}
  {2013}{\natexlab{a}})}\BibitemShut {NoStop}%
\bibitem [{\citenamefont {Dean}(2015)}]{Dean2015}%
  \BibitemOpen
  \bibfield  {author} {\bibinfo {author} {\bibfnamefont {M.~P.~M.}\
  \bibnamefont {Dean}},\ }\href {\doibase
  http://dx.doi.org/10.1016/j.jmmm.2014.03.057} {\bibfield  {journal} {\bibinfo
   {journal} {Journal of Magnetism and Magnetic Materials}\ }\textbf {\bibinfo
  {volume} {376}},\ \bibinfo {pages} {3 } (\bibinfo {year} {2015})}\BibitemShut
  {NoStop}%
\bibitem [{\citenamefont {{Lebert}}\ \emph {et~al.}(2016)\citenamefont
  {{Lebert}}, \citenamefont {{Dean}}, \citenamefont {{Nicolaou}}, \citenamefont
  {{Pelliciari}}, \citenamefont {{Dantz}}, \citenamefont {{Schmitt}},
  \citenamefont {{Yu}}, \citenamefont {{Azuma}}, \citenamefont {{Castellan}},
  \citenamefont {{Miao}}, \citenamefont {{Gauzzi}}, \citenamefont
  {{Baptiste}},\ and\ \citenamefont {{d'Astuto}}}]{Lebert2016}%
  \BibitemOpen
  \bibfield  {author} {\bibinfo {author} {\bibfnamefont {B.~W.}\ \bibnamefont
  {{Lebert}}}, \bibinfo {author} {\bibfnamefont {M.~P.~M.}\ \bibnamefont
  {{Dean}}}, \bibinfo {author} {\bibfnamefont {A.}~\bibnamefont {{Nicolaou}}},
  \bibinfo {author} {\bibfnamefont {J.}~\bibnamefont {{Pelliciari}}}, \bibinfo
  {author} {\bibfnamefont {M.}~\bibnamefont {{Dantz}}}, \bibinfo {author}
  {\bibfnamefont {T.}~\bibnamefont {{Schmitt}}}, \bibinfo {author}
  {\bibfnamefont {R.}~\bibnamefont {{Yu}}}, \bibinfo {author} {\bibfnamefont
  {M.}~\bibnamefont {{Azuma}}}, \bibinfo {author} {\bibfnamefont
  {J.}~\bibnamefont {{Castellan}}}, \bibinfo {author} {\bibfnamefont
  {H.}~\bibnamefont {{Miao}}}, \bibinfo {author} {\bibfnamefont
  {A.}~\bibnamefont {{Gauzzi}}}, \bibinfo {author} {\bibfnamefont
  {B.}~\bibnamefont {{Baptiste}}}, \ and\ \bibinfo {author} {\bibfnamefont
  {M.}~\bibnamefont {{d'Astuto}}},\ }\href@noop {} {\bibfield  {journal}
  {\bibinfo  {journal} {ArXiv e-prints}\ } (\bibinfo {year} {2016})},\ \Eprint
  {http://arxiv.org/abs/1610.08383} {arXiv:1610.08383 [cond-mat.str-el]}
  \BibitemShut {NoStop}%
\bibitem [{\citenamefont {Dantz}\ \emph {et~al.}(2016)\citenamefont {Dantz},
  \citenamefont {Pelliciari}, \citenamefont {Samal}, \citenamefont {Bisogni},
  \citenamefont {Huang}, \citenamefont {Olalde-Velasco}, \citenamefont
  {Strocov}, \citenamefont {Koster},\ and\ \citenamefont
  {Schmitt}}]{Dantz2016}%
  \BibitemOpen
  \bibfield  {author} {\bibinfo {author} {\bibfnamefont {M.}~\bibnamefont
  {Dantz}}, \bibinfo {author} {\bibfnamefont {J.}~\bibnamefont {Pelliciari}},
  \bibinfo {author} {\bibfnamefont {D.}~\bibnamefont {Samal}}, \bibinfo
  {author} {\bibfnamefont {V.}~\bibnamefont {Bisogni}}, \bibinfo {author}
  {\bibfnamefont {Y.}~\bibnamefont {Huang}}, \bibinfo {author} {\bibfnamefont
  {P.}~\bibnamefont {Olalde-Velasco}}, \bibinfo {author} {\bibfnamefont
  {V.}~\bibnamefont {Strocov}}, \bibinfo {author} {\bibfnamefont
  {G.}~\bibnamefont {Koster}}, \ and\ \bibinfo {author} {\bibfnamefont
  {T.}~\bibnamefont {Schmitt}},\ }\href@noop {} {\bibfield  {journal} {\bibinfo
   {journal} {Scientific Reports}\ }\textbf {\bibinfo {volume} {6}} (\bibinfo
  {year} {2016})}\BibitemShut {NoStop}%
\bibitem [{\citenamefont {{Bourges}}\ \emph {et~al.}(2000)\citenamefont
  {{Bourges}}, \citenamefont {{Sidis}}, \citenamefont {{Fong}}, \citenamefont
  {{Regnault}}, \citenamefont {{Bossy}}, \citenamefont {{Ivanov}},\ and\
  \citenamefont {{Keimer}}}]{Bourges2000}%
  \BibitemOpen
  \bibfield  {author} {\bibinfo {author} {\bibfnamefont {P.}~\bibnamefont
  {{Bourges}}}, \bibinfo {author} {\bibfnamefont {Y.}~\bibnamefont {{Sidis}}},
  \bibinfo {author} {\bibfnamefont {H.~F.}\ \bibnamefont {{Fong}}}, \bibinfo
  {author} {\bibfnamefont {L.~P.}\ \bibnamefont {{Regnault}}}, \bibinfo
  {author} {\bibfnamefont {J.}~\bibnamefont {{Bossy}}}, \bibinfo {author}
  {\bibfnamefont {A.}~\bibnamefont {{Ivanov}}}, \ and\ \bibinfo {author}
  {\bibfnamefont {B.}~\bibnamefont {{Keimer}}},\ }\href {\doibase
  10.1126/science.288.5469.1234} {\bibfield  {journal} {\bibinfo  {journal}
  {Science}\ }\textbf {\bibinfo {volume} {288}},\ \bibinfo {pages} {1234}
  (\bibinfo {year} {2000})}\BibitemShut {NoStop}%
\bibitem [{\citenamefont {{Tranquada}}\ \emph {et~al.}(2004)\citenamefont
  {{Tranquada}}, \citenamefont {{Woo}}, \citenamefont {{Perring}},
  \citenamefont {{Goka}}, \citenamefont {{Gu}}, \citenamefont {{Xu}},
  \citenamefont {{Fujita}},\ and\ \citenamefont {{Yamada}}}]{Tranquada2004}%
  \BibitemOpen
  \bibfield  {author} {\bibinfo {author} {\bibfnamefont {J.~M.}\ \bibnamefont
  {{Tranquada}}}, \bibinfo {author} {\bibfnamefont {H.}~\bibnamefont {{Woo}}},
  \bibinfo {author} {\bibfnamefont {T.~G.}\ \bibnamefont {{Perring}}}, \bibinfo
  {author} {\bibfnamefont {H.}~\bibnamefont {{Goka}}}, \bibinfo {author}
  {\bibfnamefont {G.~D.}\ \bibnamefont {{Gu}}}, \bibinfo {author}
  {\bibfnamefont {G.}~\bibnamefont {{Xu}}}, \bibinfo {author} {\bibfnamefont
  {M.}~\bibnamefont {{Fujita}}}, \ and\ \bibinfo {author} {\bibfnamefont
  {K.}~\bibnamefont {{Yamada}}},\ }\href {\doibase 10.1038/nature02574}
  {\bibfield  {journal} {\bibinfo  {journal} {Nature}\ }\textbf {\bibinfo
  {volume} {429}},\ \bibinfo {pages} {534} (\bibinfo {year}
  {2004})}\BibitemShut {NoStop}%
\bibitem [{\citenamefont {Hayden}\ \emph {et~al.}(2004)\citenamefont {Hayden},
  \citenamefont {Mook}, \citenamefont {Dai}, \citenamefont {Perring},\ and\
  \citenamefont {Dogan}}]{Hayden2004}%
  \BibitemOpen
  \bibfield  {author} {\bibinfo {author} {\bibfnamefont {S.~M.}\ \bibnamefont
  {Hayden}}, \bibinfo {author} {\bibfnamefont {H.~A.}\ \bibnamefont {Mook}},
  \bibinfo {author} {\bibfnamefont {P.}~\bibnamefont {Dai}}, \bibinfo {author}
  {\bibfnamefont {T.~G.}\ \bibnamefont {Perring}}, \ and\ \bibinfo {author}
  {\bibfnamefont {F.}~\bibnamefont {Dogan}},\ }\href@noop {} {\bibfield
  {journal} {\bibinfo  {journal} {Nature}\ }\textbf {\bibinfo {volume} {429}},\
  \bibinfo {pages} {531} (\bibinfo {year} {2004})}\BibitemShut {NoStop}%
\bibitem [{\citenamefont {Woo}\ \emph {et~al.}(2006)\citenamefont {Woo},
  \citenamefont {Dai}, \citenamefont {Hayden}, \citenamefont {Mook},
  \citenamefont {Dahm}, \citenamefont {Scalapino}, \citenamefont {Perring},\
  and\ \citenamefont {Dogan}}]{Woo2006}%
  \BibitemOpen
  \bibfield  {author} {\bibinfo {author} {\bibfnamefont {H.}~\bibnamefont
  {Woo}}, \bibinfo {author} {\bibfnamefont {P.}~\bibnamefont {Dai}}, \bibinfo
  {author} {\bibfnamefont {S.~M.}\ \bibnamefont {Hayden}}, \bibinfo {author}
  {\bibfnamefont {H.~A.}\ \bibnamefont {Mook}}, \bibinfo {author}
  {\bibfnamefont {T.}~\bibnamefont {Dahm}}, \bibinfo {author} {\bibfnamefont
  {D.~J.}\ \bibnamefont {Scalapino}}, \bibinfo {author} {\bibfnamefont {T.~G.}\
  \bibnamefont {Perring}}, \ and\ \bibinfo {author} {\bibfnamefont
  {F.}~\bibnamefont {Dogan}},\ }\href@noop {} {\bibfield  {journal} {\bibinfo
  {journal} {Nature Physics}\ }\textbf {\bibinfo {volume} {2}},\ \bibinfo
  {pages} {600} (\bibinfo {year} {2006})}\BibitemShut {NoStop}%
\bibitem [{\citenamefont {Lipscombe}\ \emph {et~al.}(2007)\citenamefont
  {Lipscombe}, \citenamefont {Hayden}, \citenamefont {Vignolle}, \citenamefont
  {McMorrow},\ and\ \citenamefont {Perring}}]{Lipscombe2007}%
  \BibitemOpen
  \bibfield  {author} {\bibinfo {author} {\bibfnamefont {O.~J.}\ \bibnamefont
  {Lipscombe}}, \bibinfo {author} {\bibfnamefont {S.~M.}\ \bibnamefont
  {Hayden}}, \bibinfo {author} {\bibfnamefont {B.}~\bibnamefont {Vignolle}},
  \bibinfo {author} {\bibfnamefont {D.~F.}\ \bibnamefont {McMorrow}}, \ and\
  \bibinfo {author} {\bibfnamefont {T.~G.}\ \bibnamefont {Perring}},\ }\href
  {\doibase 10.1103/PhysRevLett.99.067002} {\bibfield  {journal} {\bibinfo
  {journal} {Phys. Rev. Lett.}\ }\textbf {\bibinfo {volume} {99}},\ \bibinfo
  {pages} {067002} (\bibinfo {year} {2007})}\BibitemShut {NoStop}%
\bibitem [{\citenamefont {Vignolle}\ \emph {et~al.}(2007)\citenamefont
  {Vignolle}, \citenamefont {Hayden}, \citenamefont {McMorrow}, \citenamefont
  {Ronnow}, \citenamefont {Lake}, \citenamefont {Frost},\ and\ \citenamefont
  {Perring}}]{Vignolle2007}%
  \BibitemOpen
  \bibfield  {author} {\bibinfo {author} {\bibfnamefont {B.}~\bibnamefont
  {Vignolle}}, \bibinfo {author} {\bibfnamefont {S.~M.}\ \bibnamefont
  {Hayden}}, \bibinfo {author} {\bibfnamefont {D.~F.}\ \bibnamefont
  {McMorrow}}, \bibinfo {author} {\bibfnamefont {H.~M.}\ \bibnamefont
  {Ronnow}}, \bibinfo {author} {\bibfnamefont {B.}~\bibnamefont {Lake}},
  \bibinfo {author} {\bibfnamefont {C.~D.}\ \bibnamefont {Frost}}, \ and\
  \bibinfo {author} {\bibfnamefont {T.~G.}\ \bibnamefont {Perring}},\
  }\href@noop {} {\bibfield  {journal} {\bibinfo  {journal} {Nature Phys.}\
  }\textbf {\bibinfo {volume} {3}},\ \bibinfo {pages} {163} (\bibinfo {year}
  {2007})}\BibitemShut {NoStop}%
\bibitem [{\citenamefont {Le~Tacon}\ \emph {et~al.}(2011)\citenamefont
  {Le~Tacon}, \citenamefont {Ghiringhelli}, \citenamefont {Chaloupka},
  \citenamefont {Sala}, \citenamefont {Hinkov}, \citenamefont {Haverkort},
  \citenamefont {Minola}, \citenamefont {Bakr}, \citenamefont {Zhou},
  \citenamefont {Blanco-Canosa}, \citenamefont {Monney}, \citenamefont {Song},
  \citenamefont {Sun}, \citenamefont {Lin}, \citenamefont {Luca},
  \citenamefont {Salluzzo}, \citenamefont {Khaliullin}, \citenamefont
  {Schmitt}, \citenamefont {Braicovich},\ and\ \citenamefont
  {Keimer}}]{LeTacon2011}%
  \BibitemOpen
  \bibfield  {author} {\bibinfo {author} {\bibfnamefont {M.}~\bibnamefont
  {Le~Tacon}}, \bibinfo {author} {\bibfnamefont {G.}~\bibnamefont
  {Ghiringhelli}}, \bibinfo {author} {\bibfnamefont {J.}~\bibnamefont
  {Chaloupka}}, \bibinfo {author} {\bibfnamefont {M.~M.}\ \bibnamefont {Sala}},
  \bibinfo {author} {\bibfnamefont {V.}~\bibnamefont {Hinkov}}, \bibinfo
  {author} {\bibfnamefont {M.~W.}\ \bibnamefont {Haverkort}}, \bibinfo {author}
  {\bibfnamefont {M.}~\bibnamefont {Minola}}, \bibinfo {author} {\bibfnamefont
  {M.}~\bibnamefont {Bakr}}, \bibinfo {author} {\bibfnamefont {K.~J.}\
  \bibnamefont {Zhou}}, \bibinfo {author} {\bibfnamefont {S.}~\bibnamefont
  {Blanco-Canosa}}, \bibinfo {author} {\bibfnamefont {C.}~\bibnamefont
  {Monney}}, \bibinfo {author} {\bibfnamefont {Y.~T.}\ \bibnamefont {Song}},
  \bibinfo {author} {\bibfnamefont {G.~L.}\ \bibnamefont {Sun}}, \bibinfo
  {author} {\bibfnamefont {C.~T.}\ \bibnamefont {Lin}}, \bibinfo {author}
  {\bibfnamefont {G.~M. de}\ \bibnamefont {Luca}}, \bibinfo {author}
  {\bibfnamefont {M.}~\bibnamefont {Salluzzo}}, \bibinfo {author}
  {\bibfnamefont {G.}~\bibnamefont {Khaliullin}}, \bibinfo {author}
  {\bibfnamefont {T.}~\bibnamefont {Schmitt}}, \bibinfo {author} {\bibfnamefont
  {L.}~\bibnamefont {Braicovich}}, \ and\ \bibinfo {author} {\bibfnamefont
  {B.}~\bibnamefont {Keimer}},\ }\href {http://dx.doi.org/10.1038/nphys2041}
  {\bibfield  {journal} {\bibinfo  {journal} {Nat. Phys.}\ }\textbf {\bibinfo
  {volume} {7}},\ \bibinfo {pages} {725} (\bibinfo {year} {2011})}\BibitemShut
  {NoStop}%
\bibitem [{\citenamefont {Bisogni}\ \emph {et~al.}(2012)\citenamefont
  {Bisogni}, \citenamefont {Moretti~Sala}, \citenamefont {Bendounan},
  \citenamefont {Brookes}, \citenamefont {Ghiringhelli},\ and\ \citenamefont
  {Braicovich}}]{Bisogni2012}%
  \BibitemOpen
  \bibfield  {author} {\bibinfo {author} {\bibfnamefont {V.}~\bibnamefont
  {Bisogni}}, \bibinfo {author} {\bibfnamefont {M.}~\bibnamefont
  {Moretti~Sala}}, \bibinfo {author} {\bibfnamefont {A.}~\bibnamefont
  {Bendounan}}, \bibinfo {author} {\bibfnamefont {N.~B.}\ \bibnamefont
  {Brookes}}, \bibinfo {author} {\bibfnamefont {G.}~\bibnamefont
  {Ghiringhelli}}, \ and\ \bibinfo {author} {\bibfnamefont {L.}~\bibnamefont
  {Braicovich}},\ }\href {\doibase 10.1103/PhysRevB.85.214528} {\bibfield
  {journal} {\bibinfo  {journal} {Phys. Rev. B}\ }\textbf {\bibinfo {volume}
  {85}},\ \bibinfo {pages} {214528} (\bibinfo {year} {2012})}\BibitemShut
  {NoStop}%
\bibitem [{\citenamefont {Dean}\ \emph
  {et~al.}(2013{\natexlab{b}})\citenamefont {Dean}, \citenamefont {James},
  \citenamefont {Springell}, \citenamefont {Liu}, \citenamefont {Monney},
  \citenamefont {Zhou}, \citenamefont {Konik}, \citenamefont {Wen},
  \citenamefont {Xu}, \citenamefont {Gu}, \citenamefont {Strocov},
  \citenamefont {Schmitt},\ and\ \citenamefont {Hill}}]{DeanBSCCO2013}%
  \BibitemOpen
  \bibfield  {author} {\bibinfo {author} {\bibfnamefont {M.~P.~M.}\
  \bibnamefont {Dean}}, \bibinfo {author} {\bibfnamefont {A.~J.~A.}\
  \bibnamefont {James}}, \bibinfo {author} {\bibfnamefont {R.~S.}\ \bibnamefont
  {Springell}}, \bibinfo {author} {\bibfnamefont {X.}~\bibnamefont {Liu}},
  \bibinfo {author} {\bibfnamefont {C.}~\bibnamefont {Monney}}, \bibinfo
  {author} {\bibfnamefont {K.~J.}\ \bibnamefont {Zhou}}, \bibinfo {author}
  {\bibfnamefont {R.~M.}\ \bibnamefont {Konik}}, \bibinfo {author}
  {\bibfnamefont {J.~S.}\ \bibnamefont {Wen}}, \bibinfo {author} {\bibfnamefont
  {Z.~J.}\ \bibnamefont {Xu}}, \bibinfo {author} {\bibfnamefont {G.~D.}\
  \bibnamefont {Gu}}, \bibinfo {author} {\bibfnamefont {V.~N.}\ \bibnamefont
  {Strocov}}, \bibinfo {author} {\bibfnamefont {T.}~\bibnamefont {Schmitt}}, \
  and\ \bibinfo {author} {\bibfnamefont {J.~P.}\ \bibnamefont {Hill}},\ }\href
  {\doibase 10.1103/PhysRevLett.110.147001} {\bibfield  {journal} {\bibinfo
  {journal} {Phys. Rev. Lett.}\ }\textbf {\bibinfo {volume} {110}},\ \bibinfo
  {pages} {147001} (\bibinfo {year} {2013}{\natexlab{b}})}\BibitemShut
  {NoStop}%
\bibitem [{\citenamefont {Dean}\ \emph {et~al.}(2014)\citenamefont {Dean},
  \citenamefont {James}, \citenamefont {Walters}, \citenamefont {Bisogni},
  \citenamefont {Jarrige}, \citenamefont {H\"ucker}, \citenamefont {Giannini},
  \citenamefont {Fujita}, \citenamefont {Pelliciari}, \citenamefont {Huang},
  \citenamefont {Konik}, \citenamefont {Schmitt},\ and\ \citenamefont
  {Hill}}]{DeanBSCCO2014}%
  \BibitemOpen
  \bibfield  {author} {\bibinfo {author} {\bibfnamefont {M.~P.~M.}\
  \bibnamefont {Dean}}, \bibinfo {author} {\bibfnamefont {A.~J.~A.}\
  \bibnamefont {James}}, \bibinfo {author} {\bibfnamefont {A.~C.}\ \bibnamefont
  {Walters}}, \bibinfo {author} {\bibfnamefont {V.}~\bibnamefont {Bisogni}},
  \bibinfo {author} {\bibfnamefont {I.}~\bibnamefont {Jarrige}}, \bibinfo
  {author} {\bibfnamefont {M.}~\bibnamefont {H\"ucker}}, \bibinfo {author}
  {\bibfnamefont {E.}~\bibnamefont {Giannini}}, \bibinfo {author}
  {\bibfnamefont {M.}~\bibnamefont {Fujita}}, \bibinfo {author} {\bibfnamefont
  {J.}~\bibnamefont {Pelliciari}}, \bibinfo {author} {\bibfnamefont {Y.~B.}\
  \bibnamefont {Huang}}, \bibinfo {author} {\bibfnamefont {R.~M.}\ \bibnamefont
  {Konik}}, \bibinfo {author} {\bibfnamefont {T.}~\bibnamefont {Schmitt}}, \
  and\ \bibinfo {author} {\bibfnamefont {J.~P.}\ \bibnamefont {Hill}},\ }\href
  {\doibase 10.1103/PhysRevB.90.220506} {\bibfield  {journal} {\bibinfo
  {journal} {Phys. Rev. B}\ }\textbf {\bibinfo {volume} {90}},\ \bibinfo
  {pages} {220506} (\bibinfo {year} {2014})}\BibitemShut {NoStop}%
\bibitem [{\citenamefont {Guarise}\ \emph {et~al.}(2014)\citenamefont
  {Guarise}, \citenamefont {Dalla~Piazza}, \citenamefont {Berger},
  \citenamefont {Giannini}, \citenamefont {Schmitt}, \citenamefont {R{\o}nnow},
  \citenamefont {Sawatzky}, \citenamefont {Van Den~Brink}, \citenamefont
  {Altenfeld}, \citenamefont {Eremin} \emph {et~al.}}]{Guarise2014}%
  \BibitemOpen
  \bibfield  {author} {\bibinfo {author} {\bibfnamefont {M.}~\bibnamefont
  {Guarise}}, \bibinfo {author} {\bibfnamefont {B.}~\bibnamefont
  {Dalla~Piazza}}, \bibinfo {author} {\bibfnamefont {H.}~\bibnamefont
  {Berger}}, \bibinfo {author} {\bibfnamefont {E.}~\bibnamefont {Giannini}},
  \bibinfo {author} {\bibfnamefont {T.}~\bibnamefont {Schmitt}}, \bibinfo
  {author} {\bibfnamefont {H.~M.}\ \bibnamefont {R{\o}nnow}}, \bibinfo {author}
  {\bibfnamefont {G.}~\bibnamefont {Sawatzky}}, \bibinfo {author}
  {\bibfnamefont {J.}~\bibnamefont {Van Den~Brink}}, \bibinfo {author}
  {\bibfnamefont {D.}~\bibnamefont {Altenfeld}}, \bibinfo {author}
  {\bibfnamefont {I.}~\bibnamefont {Eremin}},  \emph {et~al.},\ }\href@noop {}
  {\bibfield  {journal} {\bibinfo  {journal} {Nature communications}\ }\textbf
  {\bibinfo {volume} {5}},\ \bibinfo {pages} {5760} (\bibinfo {year}
  {2014})}\BibitemShut {NoStop}%
\bibitem [{\citenamefont {Wakimoto}\ \emph {et~al.}(2015)\citenamefont
  {Wakimoto}, \citenamefont {Ishii}, \citenamefont {Kimura}, \citenamefont
  {Fujita}, \citenamefont {Dellea}, \citenamefont {Kummer}, \citenamefont
  {Braicovich}, \citenamefont {Ghiringhelli}, \citenamefont {Debeer-Schmitt},\
  and\ \citenamefont {Granroth}}]{Wakimoto2015}%
  \BibitemOpen
  \bibfield  {author} {\bibinfo {author} {\bibfnamefont {S.}~\bibnamefont
  {Wakimoto}}, \bibinfo {author} {\bibfnamefont {K.}~\bibnamefont {Ishii}},
  \bibinfo {author} {\bibfnamefont {H.}~\bibnamefont {Kimura}}, \bibinfo
  {author} {\bibfnamefont {M.}~\bibnamefont {Fujita}}, \bibinfo {author}
  {\bibfnamefont {G.}~\bibnamefont {Dellea}}, \bibinfo {author} {\bibfnamefont
  {K.}~\bibnamefont {Kummer}}, \bibinfo {author} {\bibfnamefont
  {L.}~\bibnamefont {Braicovich}}, \bibinfo {author} {\bibfnamefont
  {G.}~\bibnamefont {Ghiringhelli}}, \bibinfo {author} {\bibfnamefont {L.~M.}\
  \bibnamefont {Debeer-Schmitt}}, \ and\ \bibinfo {author} {\bibfnamefont
  {G.~E.}\ \bibnamefont {Granroth}},\ }\href {\doibase
  10.1103/PhysRevB.91.184513} {\bibfield  {journal} {\bibinfo  {journal} {Phys.
  Rev. B}\ }\textbf {\bibinfo {volume} {91}},\ \bibinfo {pages} {184513}
  (\bibinfo {year} {2015})}\BibitemShut {NoStop}%
\bibitem [{\citenamefont {Ellis}\ \emph {et~al.}(2015)\citenamefont {Ellis},
  \citenamefont {Huang}, \citenamefont {Olalde-Velasco}, \citenamefont {Dantz},
  \citenamefont {Pelliciari}, \citenamefont {Drachuck}, \citenamefont {Ofer},
  \citenamefont {Bazalitsky}, \citenamefont {Berger}, \citenamefont {Schmitt},\
  and\ \citenamefont {Keren}}]{Ellis2015}%
  \BibitemOpen
  \bibfield  {author} {\bibinfo {author} {\bibfnamefont {D.~S.}\ \bibnamefont
  {Ellis}}, \bibinfo {author} {\bibfnamefont {Y.-B.}\ \bibnamefont {Huang}},
  \bibinfo {author} {\bibfnamefont {P.}~\bibnamefont {Olalde-Velasco}},
  \bibinfo {author} {\bibfnamefont {M.}~\bibnamefont {Dantz}}, \bibinfo
  {author} {\bibfnamefont {J.}~\bibnamefont {Pelliciari}}, \bibinfo {author}
  {\bibfnamefont {G.}~\bibnamefont {Drachuck}}, \bibinfo {author}
  {\bibfnamefont {R.}~\bibnamefont {Ofer}}, \bibinfo {author} {\bibfnamefont
  {G.}~\bibnamefont {Bazalitsky}}, \bibinfo {author} {\bibfnamefont
  {J.}~\bibnamefont {Berger}}, \bibinfo {author} {\bibfnamefont
  {T.}~\bibnamefont {Schmitt}}, \ and\ \bibinfo {author} {\bibfnamefont
  {A.}~\bibnamefont {Keren}},\ }\href {\doibase 10.1103/PhysRevB.92.104507}
  {\bibfield  {journal} {\bibinfo  {journal} {Phys. Rev. B}\ }\textbf {\bibinfo
  {volume} {92}},\ \bibinfo {pages} {104507} (\bibinfo {year}
  {2015})}\BibitemShut {NoStop}%
\bibitem [{\citenamefont {Huang}\ \emph {et~al.}(2016)\citenamefont {Huang},
  \citenamefont {Jia}, \citenamefont {Chen}, \citenamefont {Wohlfeld},
  \citenamefont {Moritz}, \citenamefont {Devereaux}, \citenamefont {Wu},
  \citenamefont {Okamoto}, \citenamefont {Lee}, \citenamefont {Hashimoto} \emph
  {et~al.}}]{Huang2016}%
  \BibitemOpen
  \bibfield  {author} {\bibinfo {author} {\bibfnamefont {H.}~\bibnamefont
  {Huang}}, \bibinfo {author} {\bibfnamefont {C.}~\bibnamefont {Jia}}, \bibinfo
  {author} {\bibfnamefont {Z.}~\bibnamefont {Chen}}, \bibinfo {author}
  {\bibfnamefont {K.}~\bibnamefont {Wohlfeld}}, \bibinfo {author}
  {\bibfnamefont {B.}~\bibnamefont {Moritz}}, \bibinfo {author} {\bibfnamefont
  {T.}~\bibnamefont {Devereaux}}, \bibinfo {author} {\bibfnamefont
  {W.}~\bibnamefont {Wu}}, \bibinfo {author} {\bibfnamefont {J.}~\bibnamefont
  {Okamoto}}, \bibinfo {author} {\bibfnamefont {W.}~\bibnamefont {Lee}},
  \bibinfo {author} {\bibfnamefont {M.}~\bibnamefont {Hashimoto}},  \emph
  {et~al.},\ }\href@noop {} {\bibfield  {journal} {\bibinfo  {journal}
  {Scientific reports}\ }\textbf {\bibinfo {volume} {6}},\ \bibinfo {pages}
  {19657} (\bibinfo {year} {2016})}\BibitemShut {NoStop}%
\bibitem [{\citenamefont {Monney}\ \emph {et~al.}(2016)\citenamefont {Monney},
  \citenamefont {Schmitt}, \citenamefont {Matt}, \citenamefont {Mesot},
  \citenamefont {Strocov}, \citenamefont {Lipscombe}, \citenamefont {Hayden},\
  and\ \citenamefont {Chang}}]{Monney2016}%
  \BibitemOpen
  \bibfield  {author} {\bibinfo {author} {\bibfnamefont {C.}~\bibnamefont
  {Monney}}, \bibinfo {author} {\bibfnamefont {T.}~\bibnamefont {Schmitt}},
  \bibinfo {author} {\bibfnamefont {C.~E.}\ \bibnamefont {Matt}}, \bibinfo
  {author} {\bibfnamefont {J.}~\bibnamefont {Mesot}}, \bibinfo {author}
  {\bibfnamefont {V.~N.}\ \bibnamefont {Strocov}}, \bibinfo {author}
  {\bibfnamefont {O.~J.}\ \bibnamefont {Lipscombe}}, \bibinfo {author}
  {\bibfnamefont {S.~M.}\ \bibnamefont {Hayden}}, \ and\ \bibinfo {author}
  {\bibfnamefont {J.}~\bibnamefont {Chang}},\ }\href {\doibase
  10.1103/PhysRevB.93.075103} {\bibfield  {journal} {\bibinfo  {journal} {Phys.
  Rev. B}\ }\textbf {\bibinfo {volume} {93}},\ \bibinfo {pages} {075103}
  (\bibinfo {year} {2016})}\BibitemShut {NoStop}%
\bibitem [{\citenamefont {Fujita}\ \emph {et~al.}(2012)\citenamefont {Fujita},
  \citenamefont {Hiraka}, \citenamefont {Matsuda}, \citenamefont {Matsuura},
  \citenamefont {Tranquada}, \citenamefont {Wakimoto}, \citenamefont {Xu},\
  and\ \citenamefont {Yamada}}]{Fujita2012NS}%
  \BibitemOpen
  \bibfield  {author} {\bibinfo {author} {\bibfnamefont {M.}~\bibnamefont
  {Fujita}}, \bibinfo {author} {\bibfnamefont {H.}~\bibnamefont {Hiraka}},
  \bibinfo {author} {\bibfnamefont {M.}~\bibnamefont {Matsuda}}, \bibinfo
  {author} {\bibfnamefont {M.}~\bibnamefont {Matsuura}}, \bibinfo {author}
  {\bibfnamefont {J.~M.}\ \bibnamefont {Tranquada}}, \bibinfo {author}
  {\bibfnamefont {S.}~\bibnamefont {Wakimoto}}, \bibinfo {author}
  {\bibfnamefont {G.}~\bibnamefont {Xu}}, \ and\ \bibinfo {author}
  {\bibfnamefont {K.}~\bibnamefont {Yamada}},\ }\href {\doibase
  10.1143/JPSJ.81.011007} {\bibfield  {journal} {\bibinfo  {journal} {Journal
  of the Physical Society of Japan}\ }\textbf {\bibinfo {volume} {81}},\
  \bibinfo {pages} {011007} (\bibinfo {year} {2012})}\BibitemShut {NoStop}%
\bibitem [{\citenamefont {Wakimoto}\ \emph {et~al.}(2004)\citenamefont
  {Wakimoto}, \citenamefont {Zhang}, \citenamefont {Yamada}, \citenamefont
  {Swainson}, \citenamefont {Kim},\ and\ \citenamefont
  {Birgeneau}}]{Wakimoto2004}%
  \BibitemOpen
  \bibfield  {author} {\bibinfo {author} {\bibfnamefont {S.}~\bibnamefont
  {Wakimoto}}, \bibinfo {author} {\bibfnamefont {H.}~\bibnamefont {Zhang}},
  \bibinfo {author} {\bibfnamefont {K.}~\bibnamefont {Yamada}}, \bibinfo
  {author} {\bibfnamefont {I.}~\bibnamefont {Swainson}}, \bibinfo {author}
  {\bibfnamefont {H.}~\bibnamefont {Kim}}, \ and\ \bibinfo {author}
  {\bibfnamefont {R.~J.}\ \bibnamefont {Birgeneau}},\ }\href {\doibase
  10.1103/PhysRevLett.92.217004} {\bibfield  {journal} {\bibinfo  {journal}
  {Phys. Rev. Lett.}\ }\textbf {\bibinfo {volume} {92}},\ \bibinfo {pages}
  {217004} (\bibinfo {year} {2004})}\BibitemShut {NoStop}%
\bibitem [{\citenamefont {Fujiyama}\ \emph {et~al.}(2012)\citenamefont
  {Fujiyama}, \citenamefont {Ohsumi}, \citenamefont {Komesu}, \citenamefont
  {Matsuno}, \citenamefont {Kim}, \citenamefont {Takata}, \citenamefont
  {Arima},\ and\ \citenamefont {Takagi}}]{Fujiyama2012}%
  \BibitemOpen
  \bibfield  {author} {\bibinfo {author} {\bibfnamefont {S.}~\bibnamefont
  {Fujiyama}}, \bibinfo {author} {\bibfnamefont {H.}~\bibnamefont {Ohsumi}},
  \bibinfo {author} {\bibfnamefont {T.}~\bibnamefont {Komesu}}, \bibinfo
  {author} {\bibfnamefont {J.}~\bibnamefont {Matsuno}}, \bibinfo {author}
  {\bibfnamefont {B. J.}~\bibnamefont {Kim}}, \bibinfo {author} {\bibfnamefont
  {M.}~\bibnamefont {Takata}}, \bibinfo {author} {\bibfnamefont
  {T.}~\bibnamefont {Arima}}, \ and\ \bibinfo {author} {\bibfnamefont
  {H.}~\bibnamefont {Takagi}},\ }\href@noop {} {\bibfield  {journal} {\bibinfo
  {journal} {Phys. Rev. Lett.}\ }\textbf {\bibinfo {volume} {108}},\ \bibinfo
  {pages} {247212} (\bibinfo {year} {2012})}\BibitemShut {NoStop}%
\bibitem [{\citenamefont {Ament}\ \emph {et~al.}(2011)\citenamefont {Ament},
  \citenamefont {van Veenendaal}, \citenamefont {Devereaux}, \citenamefont
  {Hill},\ and\ \citenamefont {van~den Brink}}]{Ament2011}%
  \BibitemOpen
  \bibfield  {author} {\bibinfo {author} {\bibfnamefont {L.~J.~P.}\
  \bibnamefont {Ament}}, \bibinfo {author} {\bibfnamefont {M.}~\bibnamefont
  {van Veenendaal}}, \bibinfo {author} {\bibfnamefont {T.~P.}\ \bibnamefont
  {Devereaux}}, \bibinfo {author} {\bibfnamefont {J.~P.}\ \bibnamefont {Hill}},
  \ and\ \bibinfo {author} {\bibfnamefont {J.}~\bibnamefont {van~den Brink}},\
  }\href@noop {} {\bibfield  {journal} {\bibinfo  {journal} {Rev. Mod. Phys.}\
  }\textbf {\bibinfo {volume} {83}},\ \bibinfo {pages} {705} (\bibinfo {year}
  {2011})}\BibitemShut {NoStop}%
\bibitem [{\citenamefont {James}\ \emph {et~al.}(2012)\citenamefont {James},
  \citenamefont {Konik},\ and\ \citenamefont {Rice}}]{James2012}%
  \BibitemOpen
  \bibfield  {author} {\bibinfo {author} {\bibfnamefont {A.~J.~A.}\
  \bibnamefont {James}}, \bibinfo {author} {\bibfnamefont {R.~M.}\ \bibnamefont
  {Konik}}, \ and\ \bibinfo {author} {\bibfnamefont {T.~M.}\ \bibnamefont
  {Rice}},\ }\href {\doibase 10.1103/PhysRevB.86.100508} {\bibfield  {journal}
  {\bibinfo  {journal} {Phys. Rev. B}\ }\textbf {\bibinfo {volume} {86}},\
  \bibinfo {pages} {100508} (\bibinfo {year} {2012})}\BibitemShut {NoStop}%
\bibitem [{\citenamefont {Zeyher}\ and\ \citenamefont
  {Greco}(2013)}]{Zeyher2013}%
  \BibitemOpen
  \bibfield  {author} {\bibinfo {author} {\bibfnamefont {R.}~\bibnamefont
  {Zeyher}}\ and\ \bibinfo {author} {\bibfnamefont {A.}~\bibnamefont {Greco}},\
  }\href {\doibase 10.1103/PhysRevB.87.224511} {\bibfield  {journal} {\bibinfo
  {journal} {Phys. Rev. B}\ }\textbf {\bibinfo {volume} {87}},\ \bibinfo
  {pages} {224511} (\bibinfo {year} {2013})}\BibitemShut {NoStop}%
\bibitem [{\citenamefont {Eremin}\ \emph {et~al.}(2013)\citenamefont {Eremin},
  \citenamefont {Shigapov},\ and\ \citenamefont {Thuy}}]{Eremin2013}%
  \BibitemOpen
  \bibfield  {author} {\bibinfo {author} {\bibfnamefont {M.~V.}\ \bibnamefont
  {Eremin}}, \bibinfo {author} {\bibfnamefont {I.~M.}\ \bibnamefont
  {Shigapov}}, \ and\ \bibinfo {author} {\bibfnamefont {H.~T.~D.}\ \bibnamefont
  {Thuy}},\ }\href {http://stacks.iop.org/0953-8984/25/i=34/a=345701}
  {\bibfield  {journal} {\bibinfo  {journal} {Journal of Physics: Condensed
  Matter}\ }\textbf {\bibinfo {volume} {25}},\ \bibinfo {pages} {345701}
  (\bibinfo {year} {2013})}\BibitemShut {NoStop}%
\bibitem [{\citenamefont {Li}\ \emph {et~al.}(2016)\citenamefont {Li},
  \citenamefont {Lin},\ and\ \citenamefont {Lee}}]{Li2016}%
  \BibitemOpen
  \bibfield  {author} {\bibinfo {author} {\bibfnamefont {W.-J.}\ \bibnamefont
  {Li}}, \bibinfo {author} {\bibfnamefont {C.-J.}\ \bibnamefont {Lin}}, \ and\
  \bibinfo {author} {\bibfnamefont {T.-K.}\ \bibnamefont {Lee}},\ }\href
  {\doibase 10.1103/PhysRevB.94.075127} {\bibfield  {journal} {\bibinfo
  {journal} {Phys. Rev. B}\ }\textbf {\bibinfo {volume} {94}},\ \bibinfo
  {pages} {075127} (\bibinfo {year} {2016})}\BibitemShut {NoStop}%
\bibitem [{\citenamefont {Bo\v{z}ovi\'{c}}(2001)}]{Bozovic2001}%
  \BibitemOpen
  \bibfield  {author} {\bibinfo {author} {\bibfnamefont {I.}~\bibnamefont
  {Bo\v{z}ovi\'{c}}},\ }\href {\doibase 10.1109/77.919617} {\bibfield
  {journal} {\bibinfo  {journal} {IEEE Trans. Appl. Supercond.}\ }\textbf
  {\bibinfo {volume} {11}},\ \bibinfo {pages} {2686 } (\bibinfo {year}
  {2001})}\BibitemShut {NoStop}%
\bibitem [{\citenamefont {Ghiringhelli}\ \emph {et~al.}(2006)\citenamefont
  {Ghiringhelli}, \citenamefont {Piazzalunga}, \citenamefont {Dallera},
  \citenamefont {Trezzi}, \citenamefont {Braicovich}, \citenamefont {Schmitt},
  \citenamefont {Strocov}, \citenamefont {Betemps}, \citenamefont {Patthey},
  \citenamefont {Wang},\ and\ \citenamefont {Grioni}}]{Ghiringhelli2006}%
  \BibitemOpen
  \bibfield  {author} {\bibinfo {author} {\bibfnamefont {G.}~\bibnamefont
  {Ghiringhelli}}, \bibinfo {author} {\bibfnamefont {A.}~\bibnamefont
  {Piazzalunga}}, \bibinfo {author} {\bibfnamefont {C.}~\bibnamefont
  {Dallera}}, \bibinfo {author} {\bibfnamefont {G.}~\bibnamefont {Trezzi}},
  \bibinfo {author} {\bibfnamefont {L.}~\bibnamefont {Braicovich}}, \bibinfo
  {author} {\bibfnamefont {T.}~\bibnamefont {Schmitt}}, \bibinfo {author}
  {\bibfnamefont {V.~N.}\ \bibnamefont {Strocov}}, \bibinfo {author}
  {\bibfnamefont {R.}~\bibnamefont {Betemps}}, \bibinfo {author} {\bibfnamefont
  {L.}~\bibnamefont {Patthey}}, \bibinfo {author} {\bibfnamefont
  {X.}~\bibnamefont {Wang}}, \ and\ \bibinfo {author} {\bibfnamefont
  {M.}~\bibnamefont {Grioni}},\ }\href {\doibase 10.1063/1.2372731} {\bibfield
  {journal} {\bibinfo  {journal} {Rev. Sci. Inst.}\ }\textbf {\bibinfo {volume}
  {77}},\ \bibinfo {eid} {113108} (\bibinfo {year} {2006})}\BibitemShut
  {NoStop}%
\bibitem [{\citenamefont {Strocov}\ \emph {et~al.}(2010)\citenamefont
  {Strocov}, \citenamefont {Schmitt}, \citenamefont {Flechsig}, \citenamefont
  {Schmidt}, \citenamefont {Imhof}, \citenamefont {Chen}, \citenamefont
  {Raabe}, \citenamefont {Betemps}, \citenamefont {Zimoch}, \citenamefont
  {Krempasky}, \citenamefont {Wang}, \citenamefont {Grioni}, \citenamefont
  {Piazzalunga},\ and\ \citenamefont {Patthey}}]{Strocov2010}%
  \BibitemOpen
  \bibfield  {author} {\bibinfo {author} {\bibfnamefont {V.~N.}\ \bibnamefont
  {Strocov}}, \bibinfo {author} {\bibfnamefont {T.}~\bibnamefont {Schmitt}},
  \bibinfo {author} {\bibfnamefont {U.}~\bibnamefont {Flechsig}}, \bibinfo
  {author} {\bibfnamefont {T.}~\bibnamefont {Schmidt}}, \bibinfo {author}
  {\bibfnamefont {A.}~\bibnamefont {Imhof}}, \bibinfo {author} {\bibfnamefont
  {Q.}~\bibnamefont {Chen}}, \bibinfo {author} {\bibfnamefont {J.}~\bibnamefont
  {Raabe}}, \bibinfo {author} {\bibfnamefont {R.}~\bibnamefont {Betemps}},
  \bibinfo {author} {\bibfnamefont {D.}~\bibnamefont {Zimoch}}, \bibinfo
  {author} {\bibfnamefont {J.}~\bibnamefont {Krempasky}}, \bibinfo {author}
  {\bibfnamefont {X.}~\bibnamefont {Wang}}, \bibinfo {author} {\bibfnamefont
  {M.}~\bibnamefont {Grioni}}, \bibinfo {author} {\bibfnamefont
  {A.}~\bibnamefont {Piazzalunga}}, \ and\ \bibinfo {author} {\bibfnamefont
  {L.}~\bibnamefont {Patthey}},\ }\href {\doibase {10.1107/S0909049510019862}}
  {\bibfield  {journal} {\bibinfo  {journal} {J. Synchrotron Rad.}\ }\textbf
  {\bibinfo {volume} {17}},\ \bibinfo {pages} {631} (\bibinfo {year}
  {2010})}\BibitemShut {NoStop}%
\bibitem [{\citenamefont {Lai}\ \emph {et~al.}(2014)\citenamefont {Lai},
  \citenamefont {Fung}, \citenamefont {Wu}, \citenamefont {Huang},
  \citenamefont {Fu}, \citenamefont {Lin}, \citenamefont {Huang}, \citenamefont
  {Chiu}, \citenamefont {Wang}, \citenamefont {Huang}, \citenamefont {Tseng},
  \citenamefont {Chung}, \citenamefont {Chen},\ and\ \citenamefont
  {Huang}}]{Lai2014}%
  \BibitemOpen
  \bibfield  {author} {\bibinfo {author} {\bibfnamefont {C.~H.}\ \bibnamefont
  {Lai}}, \bibinfo {author} {\bibfnamefont {H.~S.}\ \bibnamefont {Fung}},
  \bibinfo {author} {\bibfnamefont {W.~B.}\ \bibnamefont {Wu}}, \bibinfo
  {author} {\bibfnamefont {H.~Y.}\ \bibnamefont {Huang}}, \bibinfo {author}
  {\bibfnamefont {H.~W.}\ \bibnamefont {Fu}}, \bibinfo {author} {\bibfnamefont
  {S.~W.}\ \bibnamefont {Lin}}, \bibinfo {author} {\bibfnamefont {S.~W.}\
  \bibnamefont {Huang}}, \bibinfo {author} {\bibfnamefont {C.~C.}\ \bibnamefont
  {Chiu}}, \bibinfo {author} {\bibfnamefont {D.~J.}\ \bibnamefont {Wang}},
  \bibinfo {author} {\bibfnamefont {L.~J.}\ \bibnamefont {Huang}}, \bibinfo
  {author} {\bibfnamefont {T.~C.}\ \bibnamefont {Tseng}}, \bibinfo {author}
  {\bibfnamefont {S.~C.}\ \bibnamefont {Chung}}, \bibinfo {author}
  {\bibfnamefont {C.~T.}\ \bibnamefont {Chen}}, \ and\ \bibinfo {author}
  {\bibfnamefont {D.~J.}\ \bibnamefont {Huang}},\ }\href {\doibase
  10.1107/S1600577513030877} {\bibfield  {journal} {\bibinfo  {journal}
  {Journal of Synchrotron Radiation}\ }\textbf {\bibinfo {volume} {21}},\
  \bibinfo {pages} {325} (\bibinfo {year} {2014})}\BibitemShut {NoStop}%
\bibitem [{\citenamefont {Dean}\ \emph
  {et~al.}(2013{\natexlab{c}})\citenamefont {Dean}, \citenamefont {Dellea},
  \citenamefont {Minola}, \citenamefont {Wilkins}, \citenamefont {Konik},
  \citenamefont {Gu}, \citenamefont {Le~Tacon}, \citenamefont {Brookes},
  \citenamefont {Yakhou-Harris}, \citenamefont {Kummer}, \citenamefont {Hill},
  \citenamefont {Braicovich},\ and\ \citenamefont
  {Ghiringhelli}}]{DeanLBCO2013}%
  \BibitemOpen
  \bibfield  {author} {\bibinfo {author} {\bibfnamefont {M.~P.~M.}\
  \bibnamefont {Dean}}, \bibinfo {author} {\bibfnamefont {G.}~\bibnamefont
  {Dellea}}, \bibinfo {author} {\bibfnamefont {M.}~\bibnamefont {Minola}},
  \bibinfo {author} {\bibfnamefont {S.~B.}\ \bibnamefont {Wilkins}}, \bibinfo
  {author} {\bibfnamefont {R.~M.}\ \bibnamefont {Konik}}, \bibinfo {author}
  {\bibfnamefont {G.~D.}\ \bibnamefont {Gu}}, \bibinfo {author} {\bibfnamefont
  {M.}~\bibnamefont {Le~Tacon}}, \bibinfo {author} {\bibfnamefont {N.~B.}\
  \bibnamefont {Brookes}}, \bibinfo {author} {\bibfnamefont {F.}~\bibnamefont
  {Yakhou-Harris}}, \bibinfo {author} {\bibfnamefont {K.}~\bibnamefont
  {Kummer}}, \bibinfo {author} {\bibfnamefont {J.~P.}\ \bibnamefont {Hill}},
  \bibinfo {author} {\bibfnamefont {L.}~\bibnamefont {Braicovich}}, \ and\
  \bibinfo {author} {\bibfnamefont {G.}~\bibnamefont {Ghiringhelli}},\ }\href
  {\doibase 10.1103/PhysRevB.88.020403} {\bibfield  {journal} {\bibinfo
  {journal} {Phys. Rev. B}\ }\textbf {\bibinfo {volume} {88}},\ \bibinfo
  {pages} {020403} (\bibinfo {year} {2013}{\natexlab{c}})}\BibitemShut
  {NoStop}%
\bibitem [{Note1()}]{Note1}%
  \BibitemOpen
  \bibinfo {note} {See the Supplemental Material of Ref.~\cite {DeanLSCO2013}
  for a explicit definition of the function}\BibitemShut {NoStop}%
\bibitem [{\citenamefont {Lamsal}\ and\ \citenamefont
  {Montfrooij}(2016)}]{Lamsal2016}%
  \BibitemOpen
  \bibfield  {author} {\bibinfo {author} {\bibfnamefont {J.}~\bibnamefont
  {Lamsal}}\ and\ \bibinfo {author} {\bibfnamefont {W.}~\bibnamefont
  {Montfrooij}},\ }\href {\doibase 10.1103/PhysRevB.93.214513} {\bibfield
  {journal} {\bibinfo  {journal} {Phys. Rev. B}\ }\textbf {\bibinfo {volume}
  {93}},\ \bibinfo {pages} {214513} (\bibinfo {year} {2016})}\BibitemShut
  {NoStop}%
\bibitem [{sup()}]{supplemental}%
  \BibitemOpen
  \href@noop {} {}\bibinfo {note} {See Supplemental Material at [URL] for
  further details of fitting and spin wave theory.}\BibitemShut {Stop}%
\bibitem [{\citenamefont {Padilla}\ \emph {et~al.}(2005)\citenamefont
  {Padilla}, \citenamefont {Dumm}, \citenamefont {Komiya}, \citenamefont
  {Ando},\ and\ \citenamefont {Basov}}]{Padilla2005}%
  \BibitemOpen
  \bibfield  {author} {\bibinfo {author} {\bibfnamefont {W.~J.}\ \bibnamefont
  {Padilla}}, \bibinfo {author} {\bibfnamefont {M.}~\bibnamefont {Dumm}},
  \bibinfo {author} {\bibfnamefont {S.}~\bibnamefont {Komiya}}, \bibinfo
  {author} {\bibfnamefont {Y.}~\bibnamefont {Ando}}, \ and\ \bibinfo {author}
  {\bibfnamefont {D.~N.}\ \bibnamefont {Basov}},\ }\href {\doibase
  10.1103/PhysRevB.72.205101} {\bibfield  {journal} {\bibinfo  {journal} {Phys.
  Rev. B}\ }\textbf {\bibinfo {volume} {72}},\ \bibinfo {pages} {205101}
  (\bibinfo {year} {2005})}\BibitemShut {NoStop}%
\bibitem [{\citenamefont {Cohen}\ \emph {et~al.}(1990)\citenamefont {Cohen},
  \citenamefont {Pickett},\ and\ \citenamefont {Krakauer}}]{Cohen1990}%
  \BibitemOpen
  \bibfield  {author} {\bibinfo {author} {\bibfnamefont {R.~E.}\ \bibnamefont
  {Cohen}}, \bibinfo {author} {\bibfnamefont {W.~E.}\ \bibnamefont {Pickett}},
  \ and\ \bibinfo {author} {\bibfnamefont {H.}~\bibnamefont {Krakauer}},\
  }\href {\doibase 10.1103/PhysRevLett.64.2575} {\bibfield  {journal} {\bibinfo
   {journal} {Phys. Rev. Lett.}\ }\textbf {\bibinfo {volume} {64}},\ \bibinfo
  {pages} {2575} (\bibinfo {year} {1990})}\BibitemShut {NoStop}%
\bibitem [{\citenamefont {Wang}\ \emph {et~al.}(1999)\citenamefont {Wang},
  \citenamefont {Yu},\ and\ \citenamefont {Krakauer}}]{Wang1999}%
  \BibitemOpen
  \bibfield  {author} {\bibinfo {author} {\bibfnamefont {C.-Z.}\ \bibnamefont
  {Wang}}, \bibinfo {author} {\bibfnamefont {R.}~\bibnamefont {Yu}}, \ and\
  \bibinfo {author} {\bibfnamefont {H.}~\bibnamefont {Krakauer}},\ }\href
  {\doibase 10.1103/PhysRevB.59.9278} {\bibfield  {journal} {\bibinfo
  {journal} {Phys. Rev. B}\ }\textbf {\bibinfo {volume} {59}},\ \bibinfo
  {pages} {9278} (\bibinfo {year} {1999})}\BibitemShut {NoStop}%
\bibitem [{\citenamefont {Minola}\ \emph {et~al.}(2015)\citenamefont {Minola},
  \citenamefont {Dellea}, \citenamefont {Gretarsson}, \citenamefont {Peng},
  \citenamefont {Lu}, \citenamefont {Porras}, \citenamefont {Loew},
  \citenamefont {Yakhou}, \citenamefont {Brookes}, \citenamefont {Huang},
  \citenamefont {Pelliciari}, \citenamefont {Schmitt}, \citenamefont
  {Ghiringhelli}, \citenamefont {Keimer}, \citenamefont {Braicovich},\ and\
  \citenamefont {Le~Tacon}}]{Minola2015}%
  \BibitemOpen
  \bibfield  {author} {\bibinfo {author} {\bibfnamefont {M.}~\bibnamefont
  {Minola}}, \bibinfo {author} {\bibfnamefont {G.}~\bibnamefont {Dellea}},
  \bibinfo {author} {\bibfnamefont {H.}~\bibnamefont {Gretarsson}}, \bibinfo
  {author} {\bibfnamefont {Y.~Y.}\ \bibnamefont {Peng}}, \bibinfo {author}
  {\bibfnamefont {Y.}~\bibnamefont {Lu}}, \bibinfo {author} {\bibfnamefont
  {J.}~\bibnamefont {Porras}}, \bibinfo {author} {\bibfnamefont
  {T.}~\bibnamefont {Loew}}, \bibinfo {author} {\bibfnamefont {F.}~\bibnamefont
  {Yakhou}}, \bibinfo {author} {\bibfnamefont {N.~B.}\ \bibnamefont {Brookes}},
  \bibinfo {author} {\bibfnamefont {Y.~B.}\ \bibnamefont {Huang}}, \bibinfo
  {author} {\bibfnamefont {J.}~\bibnamefont {Pelliciari}}, \bibinfo {author}
  {\bibfnamefont {T.}~\bibnamefont {Schmitt}}, \bibinfo {author} {\bibfnamefont
  {G.}~\bibnamefont {Ghiringhelli}}, \bibinfo {author} {\bibfnamefont
  {B.}~\bibnamefont {Keimer}}, \bibinfo {author} {\bibfnamefont
  {L.}~\bibnamefont {Braicovich}}, \ and\ \bibinfo {author} {\bibfnamefont
  {M.}~\bibnamefont {Le~Tacon}},\ }\href {\doibase
  10.1103/PhysRevLett.114.217003} {\bibfield  {journal} {\bibinfo  {journal}
  {Phys. Rev. Lett.}\ }\textbf {\bibinfo {volume} {114}},\ \bibinfo {pages}
  {217003} (\bibinfo {year} {2015})}\BibitemShut {NoStop}%
\bibitem [{\citenamefont {Jia}\ \emph {et~al.}(2014)\citenamefont {Jia},
  \citenamefont {Nowadnick}, \citenamefont {Wohlfeld}, \citenamefont {Kung},
  \citenamefont {Chen}, \citenamefont {Johnston}, \citenamefont {Tohyama},
  \citenamefont {Moritz},\ and\ \citenamefont {Devereaux}}]{Jia2014}%
  \BibitemOpen
  \bibfield  {author} {\bibinfo {author} {\bibfnamefont {C.}~\bibnamefont
  {Jia}}, \bibinfo {author} {\bibfnamefont {E.}~\bibnamefont {Nowadnick}},
  \bibinfo {author} {\bibfnamefont {K.}~\bibnamefont {Wohlfeld}}, \bibinfo
  {author} {\bibfnamefont {Y.}~\bibnamefont {Kung}}, \bibinfo {author}
  {\bibfnamefont {C.-C.}\ \bibnamefont {Chen}}, \bibinfo {author}
  {\bibfnamefont {S.}~\bibnamefont {Johnston}}, \bibinfo {author}
  {\bibfnamefont {T.}~\bibnamefont {Tohyama}}, \bibinfo {author} {\bibfnamefont
  {B.}~\bibnamefont {Moritz}}, \ and\ \bibinfo {author} {\bibfnamefont
  {T.}~\bibnamefont {Devereaux}},\ }\href@noop {} {\bibfield  {journal}
  {\bibinfo  {journal} {Nature Communications}\ }\textbf {\bibinfo {volume}
  {5}} (\bibinfo {year} {2014})}\BibitemShut {NoStop}%
\bibitem [{\citenamefont {Braicovich}\ \emph {et~al.}(2014)\citenamefont
  {Braicovich}, \citenamefont {Minola}, \citenamefont {Dellea}, \citenamefont
  {Le~Tacon}, \citenamefont {Moretti~Sala}, \citenamefont {Morawe},
  \citenamefont {Peffen}, \citenamefont {Supruangnet}, \citenamefont {Yakhou},
  \citenamefont {Ghiringhelli},\ and\ \citenamefont
  {Brookes}}]{Braicovich2014}%
  \BibitemOpen
  \bibfield  {author} {\bibinfo {author} {\bibfnamefont {L.}~\bibnamefont
  {Braicovich}}, \bibinfo {author} {\bibfnamefont {M.}~\bibnamefont {Minola}},
  \bibinfo {author} {\bibfnamefont {G.}~\bibnamefont {Dellea}}, \bibinfo
  {author} {\bibfnamefont {M.}~\bibnamefont {Le~Tacon}}, \bibinfo {author}
  {\bibfnamefont {M.}~\bibnamefont {Moretti~Sala}}, \bibinfo {author}
  {\bibfnamefont {C.}~\bibnamefont {Morawe}}, \bibinfo {author} {\bibfnamefont
  {J.-C.}\ \bibnamefont {Peffen}}, \bibinfo {author} {\bibfnamefont
  {R.}~\bibnamefont {Supruangnet}}, \bibinfo {author} {\bibfnamefont
  {F.}~\bibnamefont {Yakhou}}, \bibinfo {author} {\bibfnamefont
  {G.}~\bibnamefont {Ghiringhelli}}, \ and\ \bibinfo {author} {\bibfnamefont
  {N.~B.}\ \bibnamefont {Brookes}},\ }\href {\doibase
  http://dx.doi.org/10.1063/1.4900959} {\bibfield  {journal} {\bibinfo
  {journal} {Review of Scientific Instruments}\ }\textbf {\bibinfo {volume}
  {85}},\ \bibinfo {eid} {115104} (\bibinfo {year} {2014}),\
  http://dx.doi.org/10.1063/1.4900959}\BibitemShut {NoStop}%
\bibitem [{Note2()}]{Note2}%
  \BibitemOpen
  \bibinfo {note} {The undoped dispersion function with fixed magnetic exchange
  and a variable maximum energy transfer was used for the fitting.}\BibitemShut
  {Stop}%
\bibitem [{Note3()}]{Note3}%
  \BibitemOpen
  \bibinfo {note} {For the x = 0 case, the phonon and multi-magnon intensities
  were included with the paramagnon intensity as these were not distinguishable
  for the higher dopings.}\BibitemShut {Stop}%
\bibitem [{\citenamefont {Macridin}\ \emph {et~al.}(2007)\citenamefont
  {Macridin}, \citenamefont {Jarrell}, \citenamefont {Maier},\ and\
  \citenamefont {Scalapino}}]{Macridin2007}%
  \BibitemOpen
  \bibfield  {author} {\bibinfo {author} {\bibfnamefont {A.}~\bibnamefont
  {Macridin}}, \bibinfo {author} {\bibfnamefont {M.}~\bibnamefont {Jarrell}},
  \bibinfo {author} {\bibfnamefont {T.}~\bibnamefont {Maier}}, \ and\ \bibinfo
  {author} {\bibfnamefont {D.~J.}\ \bibnamefont {Scalapino}},\ }\href {\doibase
  10.1103/PhysRevLett.99.237001} {\bibfield  {journal} {\bibinfo  {journal}
  {Phys. Rev. Lett.}\ }\textbf {\bibinfo {volume} {99}},\ \bibinfo {pages}
  {237001} (\bibinfo {year} {2007})}\BibitemShut {NoStop}%
\bibitem [{\citenamefont {Vojta}(2009)}]{Vojta2009}%
  \BibitemOpen
  \bibfield  {author} {\bibinfo {author} {\bibfnamefont {M.}~\bibnamefont
  {Vojta}},\ }\href {\doibase 10.1080/00018730903122242} {\bibfield  {journal}
  {\bibinfo  {journal} {Advances in Physics}\ }\textbf {\bibinfo {volume}
  {58}},\ \bibinfo {pages} {699} (\bibinfo {year} {2009})}\BibitemShut
  {NoStop}%
\bibitem [{\citenamefont {Xu}\ \emph {et~al.}(2009)\citenamefont {Xu},
  \citenamefont {Gu}, \citenamefont {Hucker}, \citenamefont {Fauque},
  \citenamefont {Perring}, \citenamefont {Regnault},\ and\ \citenamefont
  {Tranquada}}]{Xu2009}%
  \BibitemOpen
  \bibfield  {author} {\bibinfo {author} {\bibfnamefont {G.}~\bibnamefont
  {Xu}}, \bibinfo {author} {\bibfnamefont {G.~D.}\ \bibnamefont {Gu}}, \bibinfo
  {author} {\bibfnamefont {M.}~\bibnamefont {Hucker}}, \bibinfo {author}
  {\bibfnamefont {B.}~\bibnamefont {Fauque}}, \bibinfo {author} {\bibfnamefont
  {T.~G.}\ \bibnamefont {Perring}}, \bibinfo {author} {\bibfnamefont {L.~P.}\
  \bibnamefont {Regnault}}, \ and\ \bibinfo {author} {\bibfnamefont {J.~M.}\
  \bibnamefont {Tranquada}},\ }\href {http://dx.doi.org/10.1038/nphys1360}
  {\bibfield  {journal} {\bibinfo  {journal} {Nat. Phys.}\ }\textbf {\bibinfo
  {volume} {5}},\ \bibinfo {pages} {642} (\bibinfo {year} {2009})}\BibitemShut
  {NoStop}%
\end{thebibliography}
\end{document}